\documentclass[prd,twocolumn,showpacs,amsmath,amssymb,letterpaper]{revtex4}
\usepackage{graphicx}
\usepackage{bm}

\begin{document}

\title{A combined analysis of short-baseline neutrino experiments
 in the (3+1) and (3+2) sterile neutrino oscillation hypotheses}

\author{M.~Sorel}
\email{sorel@nevis.columbia.edu}
\author{J.~M.~Conrad}
\email{conrad@nevis.columbia.edu}
\author{M.~H.~Shaevitz}
\email{shaevitz@nevis.columbia.edu}
\affiliation{Department of Physics, Columbia University, New York, NY 10027}
\date{\today}

\begin{abstract}
We investigate adding two sterile
 neutrinos to resolve the apparent tension existing between
 short-baseline neutrino oscillation results
 and CPT-conserving, four-neutrino oscillation models.
 For both (3+1) and (3+2) models, the level of statistical
 compatibility between the combined dataset from the null short-baseline
 experiments Bugey, CHOOZ, CCFR84, CDHS, KARMEN, and NOMAD,
 on the one hand; and the LSND dataset, on the other, is computed.
 A combined analysis of all seven short-baseline
 experiments, including LSND, is also performed, to obtain the favored regions
 in neutrino
 mass and mixing parameter space for both models. Finally,
 four statistical tests to compare the (3+1) and the
 (3+2) hypotheses are discussed. All tests show that (3+2) models fit
 the existing short-baseline data significantly better than (3+1) models.
\end{abstract}


\maketitle
\section{\label{sec:one}INTRODUCTION}
\begin{figure}[!tb]
\includegraphics*[width=3.8cm]{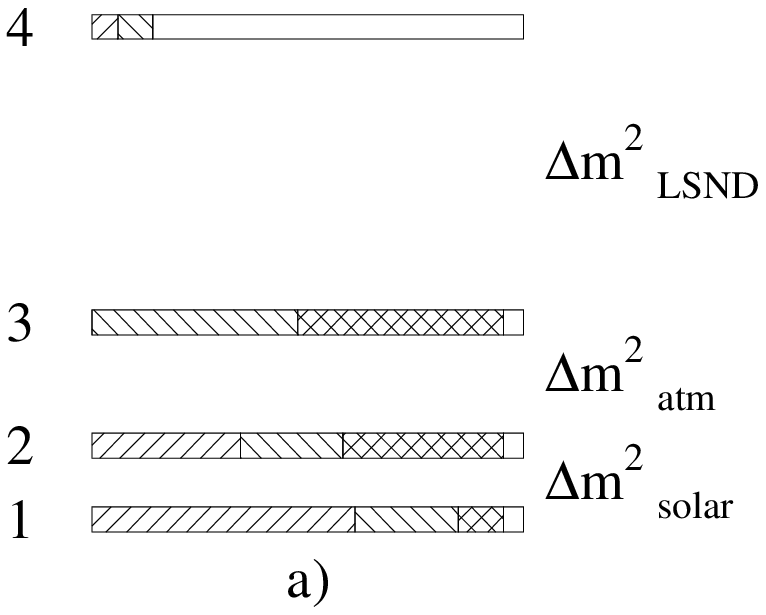} \hspace{0.4cm}
\includegraphics*[width=3.8cm]{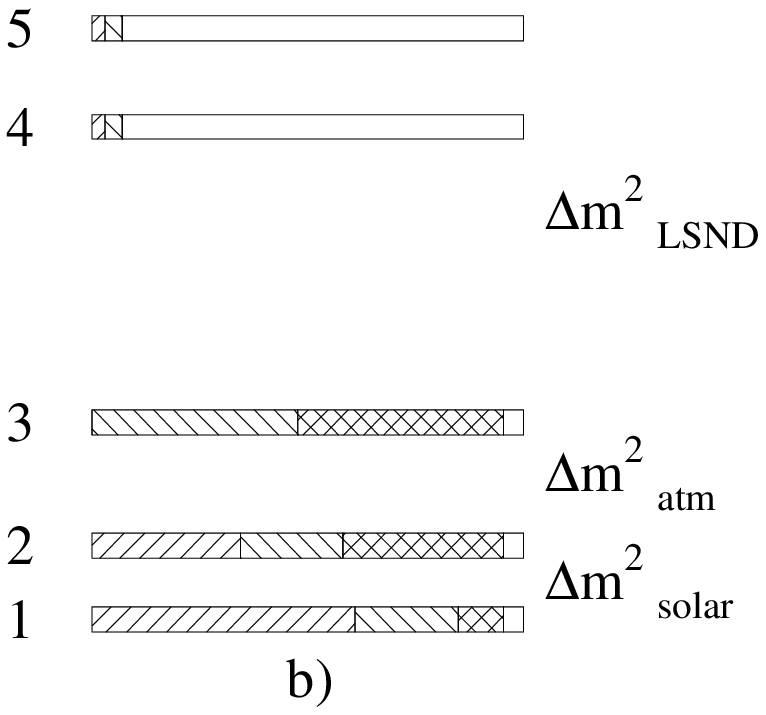}
\caption{\label{fig:fig1}Flavor content of neutrino mass eigenstates
 in (3+1) models (a), and (3+2) models (b). Neutrino masses increase from
 bottom to top. The $\nu_e$ fractions are
 indicated by right-leaning hatches, the $\nu_{\mu}$ fractions by
 left-leaning hatches, the $\nu_{\tau}$ fractions by
 crosshatches, and the $\nu_s$ fractions by no hatches. The flavor contents
 shown are schematic only.}
\end{figure}

There currently exist three experimental
 signatures for neutrino oscillations. The two signatures seen originally
 in solar and atmospheric neutrinos have
 been verified by several experiments,
 including experiments carried out with accelerator and nuclear reactor
 sources. The results on atmospheric neutrinos can be explained by
 $\nu_{\mu}$ disappearance
 due to oscillations \cite{atm1,atm2,atm3},
 while those on solar neutrinos can be explained by $\nu_e$
 oscillations \cite{sun1,sun2}.
 The third signature is $\bar{\nu}_e$ appearance in a
 $\bar{\nu}_{\mu}$ beam, observed by the short-baseline, accelerator-based
 LSND experiment at Los Alamos \cite{lsnd}. This signature is strong from
 a statistical
 point of view, being a 3.8$\sigma$ excess,
 but further confirmation by an independent experiment
 is necessary. The MiniBooNE experiment at Fermilab will be able to confirm
 or refute the LSND signature in the near future, with an experimental
 setup providing different systematics and higher statistics than
 LSND \cite{Bazarko:1999hq}. \\
\indent Taken at face value, the three experimental signatures point to three
 independent mass
 splittings. Three neutrino masses do not appear to be able
 to explain
 all of the three signatures \cite{Maltoni:2001bc,Giunti:2000wt} (see, however,
 \cite{onlythree}).
 One way to solve this puzzle is to introduce different mass
 spectra for the neutrino and antineutrino sector, thereby requiring
 CPT-violation but no extra neutrino generations
 \cite{cptv}. Another possibility
 is to add additional neutrinos
 with no standard weak couplings, often called ``sterile neutrinos''. \\
\indent In this paper we assume CPT- and CP-invariance, and
 we explore the possibility of adding one or two neutrino generations
 beyond the three active flavors assumed by the Standard Model.
 We focus on extensions of the neutrino sector
 where the addition of fourth and
 fifth mass eigenstates are responsible for the high $\Delta m^2$ LSND
 oscillations, and the three lower mass states
 explain solar and atmospheric oscillations.
 When only one sterile neutrino is added, these models are labelled
 as (3+1). The flavor content of the four neutrino mass eigenstates
 for these models is schematically shown in Fig.~\ref{fig:fig1}a.
 The (3+1) hierarchy in Fig.~\ref{fig:fig1}a is as opposed to the (2+2)
 hierarchy, where the solar and atmospheric mass splittings are separated from
 each other by the LSND $\Delta m^2$. The (2+2) models require a
 different global analysis from the one discussed in this paper.
 The simplest
 (2+2) models appear to be only
 marginally consistent with neutrino oscillations data
 \cite{Maltoni:2001bc,Maltoni:2002xd},
 even though
 more general (2+2) mass and mixing scenarios might represent a
 viable solution to explain solar, atmospheric, and LSND oscillations
 \cite{Paes:2002ah}. \\
\indent The (3+1) models are motivated by the criterion of simplicity in
 physics, introducing the most minimal extension to the Standard
 Model that explains the experimental evidence. However,
 theories invoking sterile neutrinos to explain the
 origin of neutrino masses do not necessarily require only one sterile
 neutrino. Indeed, many popular realizations of the see-saw mechanism
 introduce three right-handed neutrino fields
 \cite{Ramond:1979py,Mohapatra:1980qe,Volkas:2001zb}. In particular,
 (3+2) neutrino mass and mixing models can be obtained in several
 see-saw mechanisms \cite{threeplustwo_theory}. 
 From the phenomenological point of view, it is our opinion that
 two- and three-sterile neutrino models should also be considered
 and confronted with existing experimental results.
 In this paper, we consider the results from the short-baseline
 experiments Bugey \cite{Declais:1994su}, CCFR84 \cite{Stockdale:1984cg},
 CDHS \cite{Dydak:1983zq}, CHOOZ \cite{Apollonio:2002gd},
 KARMEN \cite{Armbruster:2002mp}, LSND \cite{lsnd}, and NOMAD
 \cite{nomad}, and
 examine how well (3+1) and (3+2) models agree with data.
 A schematic diagram for (3+2) models is shown in Fig.~\ref{fig:fig1}b.
 We do not consider (3+3) models in this paper. From our initial
 studies, we believe that the phenomenology of a (3+3) model is
 similar to a (3+2) model. \\
\indent The paper is organized as follows. In Section \ref{sec:two},
 we specify
 the neutrino oscillations formalism used in this analysis to describe
 (3+1) and (3+2), short-baseline, oscillations.
 In Section \ref{sec:three} and \ref{sec:four}, we present the results
 obtained for the
 (3+1) and (3+2) models, respectively. For both models, we first
 derive the level
 of compatibility between the null short-baseline (NSBL) experiments
 and LSND. Second, we perform a combined analysis of all
 seven short-baseline experiments (including LSND) to derive the
 preferred regions in neutrino mass and mixing parameter space.
 In Section V, we discuss four statistical tests to compare
 the (3+1) and (3+2) hypotheses. In Section VI, we briefly mention
 other experimental constraints on (3+1) and (3+2) models.
 In Appendix \ref{sec:eight}, we describe the physics and statistical
 assumptions used in the analysis to describe the short-baseline experiments.
%

\section{\label{sec:two}NEUTRINO OSCILLATIONS FORMALISM}

\indent Under the assumptions of CP- and CPT-invariance,
 the probability for a neutrino,
 produced with flavor $\alpha$ and energy $E$, to be detected as a neutrino
 of flavor $\beta$ after travelling a distance $L$, is
 \cite{Kayser:2002qs}:
\begin{equation}
P(\nu_{\alpha}\rightarrow \nu_{\beta})=
\delta_{\alpha\beta}-4\sum_{j>i}^n U_{\alpha ,j}
 U_{\beta ,j} U_{\alpha ,i} U_{\beta ,i} \sin^2 x_{ji}
\label{eq:osc_gen}
\end{equation}
\noindent where $\alpha = e,\mu ,\tau ,s$
 ($s$ being the sterile flavor); $U$ is the unitary neutrino mixing matrix;
 $x_{ji}\equiv 1.27\Delta m_{ji}^2 L/E$; $\Delta m_{ji}^2\equiv
 m_j^2-m_i^2$; and $n$ is the number of neutrino
 generations. Neglecting CP-violating phases, there are in general
 $(n-1)$ independent mass splittings, and $n^2-n-n(n-1)/2$ independent mixing
 matrix elements. The situation simplifies considerably by considering
 short-baseline (SBL) data only. In this case, it is a good
 approximation to assume $x_{21}=x_{32}=0$, and only $(n-3)$ independent
 mass splittings are present. Moreover, given the set of SBL experiments
 considered, the number of mixing matrix elements probed is only
 $2(n-3)$, as we show now for the (3+1) and (3+2) cases. \\
\indent For (3+1) models, $n$=4, and only one mass splitting
 $\Delta m^2\equiv \Delta m_{41}^2\simeq
 \Delta m^2_{42}\simeq \Delta m^2_{43}$ appears in the oscillation formula:
 this is
 sometimes referred to as to the ``quasi-two-neutrino approximation'',
 or ``one mass scale dominance'' \cite{Hagiwara:fs}. 
 Using the unitarity properties of the mixing
 matrix, we can rewrite Eq.~\ref{eq:osc_gen} for (3+1) models in a more
 convenient way:
\begin{equation}
\label{eq:osc_3+1}
P(\nu_{\alpha}\rightarrow \nu_{\beta})=
\delta_{\alpha\beta}-
4U_{\alpha 4}U_{\beta 4}
(\delta_{\alpha\beta}-U_{\alpha 4}U_{\beta 4})
\sin^2 x_{41}
\end{equation}
\noindent which depends on the mass splitting ($\Delta m_{41}^2$)
 and mixing parameters
 ($U_{\alpha 4},\ U_{\beta 4}$) of the fourth generation only. Since the
 two-neutrino approximation is satisfied in the (3+1) case, we can express
 Eq.~\ref{eq:osc_3+1} in the usual forms:
\begin{equation}
\label{eq:osc_3+1app}
P(\nu_{\alpha}\rightarrow \nu_{\beta})=
\sin^2 2\theta_{\alpha\beta}\sin^2 x_{41},\ \alpha\neq \beta
\end{equation}
\begin{equation}
\label{eq:osc_3+1dis}
P(\nu_{\alpha}\rightarrow \nu_{\alpha})=1-
\sin^2 2\theta_{\alpha\alpha}\sin^2 x_{41}
\end{equation}
\noindent where Eq.~\ref{eq:osc_3+1app} applies to an oscillation appearance
 measurement, Eq.~\ref{eq:osc_3+1dis} to a disappearance measurement. \\
\indent In this paper, we use the data from the Bugey,
 CCFR84, CDHS, CHOOZ,
 KARMEN, LSND, and NOMAD experiments. Bugey and CHOOZ data constrain
 $\bar{\nu}_e$ disappearance; CCFR84 and CDHS data
 constrain $\nu_{\mu}$ disappearance; and KARMEN, LSND, and NOMAD data constrain 
$\bar{\nu}_{\mu} \rightarrow \bar{\nu}_e$ oscillations. Therefore,
 from Eqs.~\ref{eq:osc_3+1},\ref{eq:osc_3+1app}, and \ref{eq:osc_3+1dis},
 the experiments constrain the following combinations of (3+1) mixing
 parameters: 
\begin{itemize}
\item Bugey, CHOOZ: $\sin^2 2\theta_{ee}\equiv 4U_{e4}^2(1-U_{e4}^2)$;
\item CCFR84, CDHS: $\sin^2 2\theta_{\mu\mu}\equiv 4U_{\mu 4}^2(1-U_{\mu 4}^2)$;
\item KARMEN, LSND, NOMAD:
 $\sin^2 2\theta_{\mu e}\equiv 4U_{e4}^2U_{\mu 4}^2$. 
\end{itemize}
\indent In (3+1) models, the tension between the experimental results comes
 about because Bugey,
 CHOOZ, CCFR84, CDHS, KARMEN, and NOMAD limit the two independent mixing matrix
 parameters $U_{e4}$ and $U_{\mu 4}$ to be small,
 whereas LSND demands nonzero values. \\
\indent In (3+2) models,
 we introduce two sterile neutrinos. Using
 Eq.~\ref{eq:osc_gen} and the unitarity of the mixing matrix, the (3+2)
 neutrino oscillation probability formula can be written:
\begin{widetext}
\begin{eqnarray}
P(\nu_{\alpha}\rightarrow \nu_{\beta})=
\delta_{\alpha\beta}-4[(\delta_{\alpha\beta}-U_{\alpha 4}
U_{\beta 4}-U_{\alpha 5}
U_{\beta 5})(U_{\alpha 4}U_{\beta 4}\sin^2 x_{41}+U_{\alpha 5}U_{\beta 5}
\sin^2 x_{51})+
U_{\alpha 4}U_{\alpha 5}U_{\beta 4}U_{\beta 5}
\sin^2 x_{54}] = \nonumber \\
=\delta_{\alpha\beta}+4[U_{\alpha 4}^2(U_{\beta 4}^2-\delta_{\alpha\beta})
\sin^2 x_{41}+U_{\alpha 5}^2(U_{\beta 5}^2-\delta_{\alpha\beta})\sin^2 x_{51}+
U_{\alpha 4}U_{\beta 4}U_{\alpha 5}U_{\beta 5}(\sin^2 x_{41}+\sin^2 x_{51}-
\sin^2 x_{54})]
\label{eq:osc_3+2}
\end{eqnarray}
\end{widetext}
\noindent which in our case depends on two independent mass splittings
 ($\Delta m_{41}^2,\Delta m_{51}^2$) and four independent
 mixing matrix parameters ($U_{\alpha 4},\ U_{\alpha 5},$
 with $\alpha = e,\ \mu$).
 Eq.~\ref{eq:osc_3+1} can be recovered from
 Eq.~\ref{eq:osc_3+2} by requiring
 $U_{\alpha 5}=U_{\beta 5}=0$. In (3+2) models, the
 quasi-two-neutrino-approximation is not valid, since there are three distinct
 $\Delta m^2$ values contributing in the oscillation formula:
 $\Delta m_{41}^2$, $\Delta m_{51}^2$, and $\Delta m_{54}^2$,
 and therefore three distinct
 oscillation amplitudes: $(\sin^2 2\theta_{\alpha\beta})_{41}$,
 $(\sin^2 2\theta_{\alpha\beta})_{51}$, and
 $(\sin^2 2\theta_{\alpha\beta})_{54}$. \\
\indent We now comment on the Monte Carlo method used to apply the above
 oscillation formalism to the analyses presented in this paper. 
 We require the neutrino mass splittings to be in the range
 $0.1\ \hbox{eV}^2\le \Delta m_{41}^2,\Delta m_{51}^2 \le 100\ \hbox{eV}^2$, with
 $\Delta m_{51}^2\ge \Delta m_{41}^2$. Each mass splitting range is
 analyzed over a 200 point grid, uniformly in $\log_{10}\Delta m^2$.
 In Section \ref{sec:six}, we
 briefly discuss why large mass splittings are not
 necessarily in contradiction with cosmological (and other) data.
 The values of the mixing parameters, $U_{e4}$, $U_{\mu 4}$,
 $U_{e5}$, and $U_{\mu 5}$,
 are randomly generated over a four-dimensional space satisfying the
 four requirements: $U_{ei}^2+U_{\mu i}^2\le 0.5$,
 $U_{\alpha 4}^2+U_{\alpha 5}^2\le 0.5$, where: $i=4,5$, $\alpha =e,\mu$.
 These four inequalities are introduced to account for the
 fact that large electron and muon flavor fractions in the fourth and fifth
 mass eigenstates are not allowed by solar and atmospheric neutrino data.
 In principle, since the CDHS constraint on
 $\nu_{\mu}$ disappearance
 vanishes for $\Delta m^2\simeq 0.3\ \hbox{eV}^2$, as shown in Appendix
 \ref{sec:eight},
 the upper limit on $\nu_{\mu}$ disappearance from atmospheric neutrino
 experiments above the atmospheric $\Delta m^2$ should be considered
 instead. In this paper, we do not reconstruct the
 likelihood for atmospheric data that would give the exclusion
 region for $\nu_{\mu}$ disappearance in the range
 $\Delta m^2_{atm} \ll \Delta m^2 < 0.3\ \hbox{eV}^2$. However, the effect
 that the atmospheric constraints would have on our results
 is expected to be small. For example, in Ref.~\cite{Bilenky:1999ny},
 Bilenky {\it et al.} use the atmospheric up-down asymmetry to derive
 the upper limit $U_{\mu 4}^2<0.55$ at 90\% CL for (3+1) models, which is
 satisfied by our
 initial requirements $U_{e4}^2+U_{\mu 4}^2<0.5,\
 U_{\mu 4}^2+U_{\mu 5}^2<0.5$. A more recent analysis
 \cite{Maltoni:2001mt} of atmospheric neutrino data using the full zenith angle
 distribution provides a tighter constraint on
 $\sin^2 \theta_{\mu\mu}$ than the one given in Ref. \cite{Bilenky:1999ny};
 the impact of this additional constraint on our SBL analysis is discussed
 in Sections \ref{sec:three} and \ref{sec:six}.
 Finally, from
 Eqs.~\ref{eq:osc_3+1}, it is clear that the relative sign of
 both $U_{e4}$ and $U_{\mu 4}$ cannot be inferred in (3+1)
 oscillations. Similarly, from
 Eq.~\ref{eq:osc_3+2}, the only
 physically observable relative sign between mixing parameters
 in CP-conserving (3+2) models is $sign(U_{e4}U_{\mu 4}
U_{e5}U_{\mu 5})$; therefore, this is the only sign related to
 mixing parameters that we randomly generate in the analysis. \\
\indent Throughout the paper, we make use of the Gaussian approximation
 in determining allowed regions in parameter space. In general, this means
 that the regions of quoted confidence level are the ones enclosed
 by contours of constant $\chi^2$ values, whose differences with
 respect to the best-fit $\chi^2$ value depend
 on the number of free parameters in the model \cite{James:1975dr}.
 In the text, we use the symbol $\delta$ to denote the values of the
 confidence levels derived in this way.
 As pointed out in \cite{Feldman:1997qc}, this approach should be
 considered approximate, as it may provide
 regions in parameter space of both higher and lower confidence
 than the one quoted.
 Regions of higher confidence than the quoted value may result from the
 presence
 of highly correlated parameters. Regions of lower confidence
 may result from the presence of fast oscillatory behavior
 of the oscillation probability formula, Eq.~\ref{eq:osc_gen}. \\


\section{\label{sec:three}RESULTS FOR (3+1) MODELS}

This section, like the next one on (3+2) models, consists of two
 parts. First, we quantify the statistical compatibility between
 the NSBL and LSND results, following a method described in
 \cite{Eitel:1999gt,Church:2002tc}, originally proposed to establish
 the compatibility between the LSND and KARMEN results. Second, we perform
 a combined analysis of the
 NSBL and LSND datasets, to obtain the favored regions in neutrino mass
 and mixing parameter space.

\subsection{\label{subsec:threea}Statistical compatibility between NSBL
 and LSND}

Many analyses of the NSBL experiments within (3+1)
 models have concluded that the allowed LSND region is largely
 excluded \cite{Peres:2000ic,Strumia:2002fw,Grimus:2001mn}.
 Here, we repeat this study with two purposes.  
 First, we use this study to give context to our discussion of 
 the basic model and techniques which will be expanded in later
 sections.  
 Second, we  demonstrate that our fit, which forms the basis 
 of our new results for (3+2) models, reproduces the expected (3+1)
 exclusion region.
 For a discussion of the physics and statistical assumptions used
 to describe the short-baseline experiments used in the analysis,
 the reader should refer to Appendix \ref{sec:eight}. \\
%
\begin{figure*}[!tb]
\includegraphics*[width=18.cm]{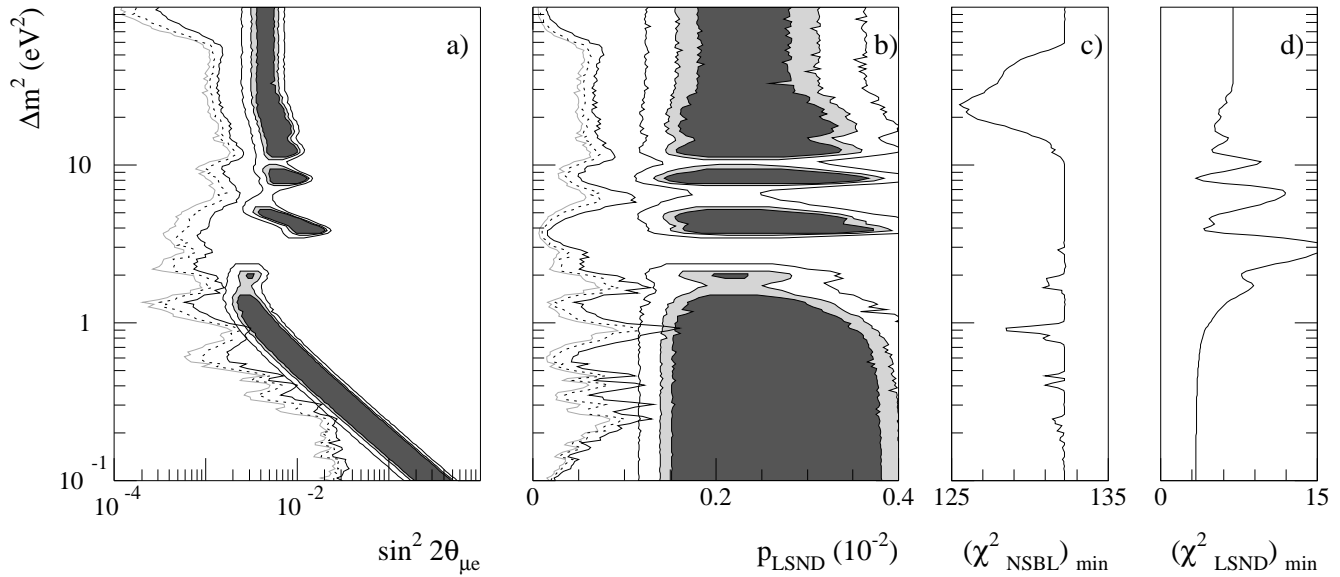}
\caption{\label{fig:fig2}Compatibility between the NSBL and LSND
 datasets in (3+1) models. Fig.~\ref{fig:fig2}a shows the 90\%
 (grey solid line),
 95\% (black dotted line), and
 99\% (black solid line) CL exclusion curves in
 $(\sin^2 2\theta_{\mu e} ,\Delta m^2)$ space for
 (3+1) models, considering the null
 short-baseline (NSBL) experiments Bugey, CCFR84, CDHS, CHOOZ, KARMEN,
 and NOMAD.
 Fig.~\ref{fig:fig2}a also shows
 the 90\%, 95\%, and 99\% CL allowed regions by our analysis of LSND data.
 Fig.~\ref{fig:fig2}b) is as Fig.~\ref{fig:fig2}a, but in
 $(p_{\hbox{\tiny LSND}},\Delta m^2)$ space, where
 $p_{\hbox{\tiny LSND}}$ is the LSND
 oscillation probability (see text for the definition).
 Fig.~\ref{fig:fig2}c) and d) show
 the minimum $\chi^2$ values as a function of $\Delta m^2$ for the NSBL
 and LSND datasets (143 and 3 d.o.f., respectively).}
\end{figure*}
\indent In this section, the NSBL and LSND datasets are analyzed
 separately, providing two independent allowed regions in
 $(\sin^2 2\theta_{\mu e},\Delta m^2)$ space.
 The level of statistical compatibility between the two datasets
 can be determined by studying to what degree the two allowed regions overlap,
 as will be quantified later in this section. \\
\indent For each randomly generated
 (3+1) model, we calculate the values for the $\chi^2$ functions
 $\chi^2_{\hbox{\tiny NSBL}}$ and $\chi^2_{\hbox{\tiny LSND}}$, where
 $\chi^2_{\hbox{\tiny NSBL}}$ is defined as:
\begin{equation}
\chi^2_{\hbox{\tiny NSBL}}\equiv
 \chi^2_{\hbox{\tiny Bugey}}+\chi^2_{\hbox{\tiny CHOOZ}}+
\chi^2_{\hbox{\tiny CCFR84}}+
\chi^2_{\hbox{\tiny CDHS}}+\chi^2_{\hbox{\tiny KARMEN}}+
\chi^2_{\hbox{\tiny NOMAD}}
\end{equation}
\indent For the analysis described in this section, the NSBL and LSND
 allowed regions are obtained using two different algorithms,
 reflecting the fact that the NSBL dataset provides upper limits on
 oscillations, while the LSND dataset points to non-null oscillations. \\ 
\indent The NSBL
 allowed regions at various confidence levels
 $\delta_{\hbox{\tiny NSBL}}$ are
 obtained via a raster scan algorithm \cite{Feldman:1997qc}. Let
$\chi^2_{\hbox{\tiny NSBL}}$
 be the $\chi^2$ value for the particular model and
 $(\chi^2_{\hbox{\tiny NSBL}})_{\hbox{\small min},\Delta m^2}$
 be the minimum $\chi^2$ for the $\Delta m^2$ value
 considered.
 For example, our quoted 95\% CL upper limit on
 $\sin^2 2\theta_{\mu e}$ is given by the maximum 
 value for the product $4U_{e4}^2U_{\mu 4}^2$ chosen among the models which
 satisfy the inequality 
 $\chi^2_{\hbox{\tiny NSBL}}-(\chi^2_{\hbox{\tiny NSBL}})_{\hbox{\small min},\Delta m^2}<5.99$. The value of 5.99 units of $\chi^2$ is chosen because there
 are two free parameters $U_{e4}$, $U_{\mu 4}$ for (3+1) models with
 fixed $\Delta m^2$. We note that even for the NSBL dataset,
 the parameters
 $U_{e4}$, $U_{\mu 4}$ can be correlated,
 since the KARMEN and NOMAD results probe a combination of the two
 parameters. \\
\indent The LSND allowed regions at various confidence levels
 $\delta_{\hbox{\tiny LSND}}$ are obtained via a global scan
 algorithm \cite{Feldman:1997qc}. For example, for
 $\delta_{\hbox{\tiny LSND}}=0.95$ we require
 $\chi^2_{\hbox{\tiny LSND}}-(\chi^2_{\hbox{\tiny LSND}})_{\hbox{\small min}}<5.99$,
 where
 $(\chi^2_{\hbox{\tiny LSND}})_{\hbox{\small min}}$ is now the global LSND
 $\chi^2$ minimum value,
 considering all possible $\Delta m^2$ values. The LSND allowed
 region is computed for two free parameters as for the NSBL case, but the
 parameters are now $\Delta m^2$ and $U_{\mu 4}U_{e4}$, as opposed to
 $U_{\mu 4}$ and $U_{e 4}$. Compared to the NSBL case,
 the number of free parameters is reduced by one because
 the LSND $\bar{\nu}_{\mu}\rightarrow \bar{\nu}_e$ search only probes
 the product $U_{\mu 4}U_{e4}$ and not the two mixing matrix elements
 individually,
 and it is increased by one
 because the allowed region is now obtained by scanning over all
 possible $\Delta m^2$ values. \\
\indent The regions allowed in $(\sin^2 2\theta_{\mu e},\Delta m^2)$
 parameter space by both the NSBL and LSND
 datasets are shown in Fig.~\ref{fig:fig2}a. The NSBL allowed regions
 shown are 2-dimensional projections of 3-dimensional allowed
 regions in $(\Delta m^2,U_{e4},U_{\mu 4})$ space.
 The NSBL results alone allow the regions to the left of
 the solid grey, dotted black, and solid black lines in the 
 Fig.~\ref{fig:fig2}a, at a confidence level
 $\delta_{\hbox{\tiny NSBL}}=0.90,\ 0.95,\ 0.99,$ 
 respectively.  In Fig.~\ref{fig:fig2}a,
 the $\delta_{LSND}=0.90,\ 0.95,\ 0.99$ CL allowed regions obtained
 by our analysis for LSND data are also shown, as dark grey shaded,
 light grey shaded, and white areas, respectively.
 We find no overlap between the two individual
 95\% CL allowed regions; on the other hand, there is overlap between
 the two 99\% CL regions. \\
\indent Fig.~\ref{fig:fig2}b shows the same (3+1) allowed regions as
 Fig.~\ref{fig:fig2}a but in the $(p_{\hbox{\tiny LSND}},\Delta m^2)$ plane,
 where $p_{\hbox{\tiny LSND}}$ is defined as the
 $\nu_{\mu}\rightarrow \nu_e$ oscillation probability
 averaged over the LSND $L/E$ distribution:
\begin{equation}
\label{eq:plsnd}
p_{LSND}\equiv \langle P(\nu_{\mu}\rightarrow \nu_{e})\rangle
\end{equation}
\noindent where $P(\nu_{\mu}\rightarrow \nu_{e})$ is given by
 Eq.~\ref{eq:osc_gen} for
 $\alpha =\mu,\ \beta =e$, and is a function of
 all the mass and mixing parameters of the oscillation model
 under consideration. This has the obvious disadvantage of being a
 quantity dependent upon the specifics of a certain experiment, as
 opposed to a universal variable such as $\sin^2 2\theta_{\mu e}=
 4U_{\mu 4}^2U_{e4}^2$. However,
 $p_{\hbox{\tiny LSND}}$ has the advantage of
 being unambiguously defined for any number of neutrino generations, and
 thus is useful in discussing (3+2) models later in this paper.
 As stated
 previously, the oscillation probability estimator
 $\sin^2 2\theta_{\mu e}=4U_{\mu 4}^2U_{e4}^2$ cannot be used
 when more than one $\Delta m^2$ value affects the oscillation
 probability, as is the case for (3+2) models. A second
 advantage of using $p_{\hbox{\tiny LSND}}$ instead of $\sin^2 2\theta_{\mu e}$
 as the oscillation probability estimator, is that the allowed values
 for $p_{\hbox{\tiny LSND}}$ inferred from the LSND result
 tend to be almost
 $\Delta m^2$-independent (see grey-shaded areas in Fig.~\ref{fig:fig2}b),
 as expected for an almost pure counting experiment such as LSND.
 The oscillation probability reported by the LSND collaboration
 \cite{lsnd} is $p_{\hbox{\tiny LSND}}=(0.264\pm 0.067\pm 0.045)\%$,
 and agrees well with our result of Fig.~\ref{fig:fig2}b. \\
\indent Fig.~\ref{fig:fig2}c shows the values for
 $(\chi^2_{\hbox{\tiny NSBL}})_{\hbox{\tiny min}}$ as a function of
 $\Delta m^2$. The number of degrees of freedom is 143.
 As discussed in
 Appendix \ref{sec:eight}, the dip in
 $(\chi^2_{\hbox{\tiny NSBL}})_{\hbox{\tiny min}}$ at
 $\Delta m^2\simeq 0.9\
 \hbox{eV}^2$ is due to Bugey data preferring $U_{e4}\neq 0$ values, while the
 minimum at $\Delta m^2\sim 10-30\ \hbox{eV}^2$ is due to CDHS (mostly) and
 CCFR84 data, preferring $U_{\mu 4}\neq 0$ values. 
 The $\chi^2$ value for no oscillations,
 $(\chi^2_{\hbox{\tiny NSBL}})_{\hbox{\tiny no osc}}=132.2$, is the largest
 $\chi^2$ value in Fig.~\ref{fig:fig2}c;
 this means that the choice of parameters $U_{e4}=U_{\mu 4}=0$
 provides the best-fit to NSBL data, for the $\Delta m^2$ values
 satisfying the condition
 $(\chi^2_{\hbox{\tiny NSBL}})_{\hbox{\tiny min}}=
(\chi^2_{\hbox{\tiny NSBL}})_{\hbox{\tiny no osc}}$.
 Note that the $\Delta m^2\simeq 0.9\ \hbox{eV}^2$, $\Delta m^2\simeq 10-30\ \hbox{eV}^2$
 dips in $\chi^2_{\hbox{\tiny NSBL}}$
 are consistent with $U_{e4}U_{\mu 4}=0$, and therefore with
 $\sin^2 2\theta_{\mu e}=p_{\hbox{\tiny LSND}}=0$, but give better 
 fits than the no-oscillations hypothesis, $U_{e4}=U_{\mu 4}=0$. In other
 words, the goodness of fit for the
 $\sin^2 2\theta_{\mu e}=p_{\hbox{\tiny LSND}}=0$
 region depends on the $\Delta m^2$ value considered. \\
\indent Similarly, Fig.~\ref{fig:fig2}d shows the values for
 $(\chi^2_{\hbox{\tiny LSND}})_{\hbox{\tiny min}}$ as a function of
 $\Delta m^2$, used to obtain the
 LSND allowed regions drawn in Figs.~\ref{fig:fig2}a, \ref{fig:fig2}b. \\
\begin{figure*}[!tb]
\includegraphics*[width=16.0cm]{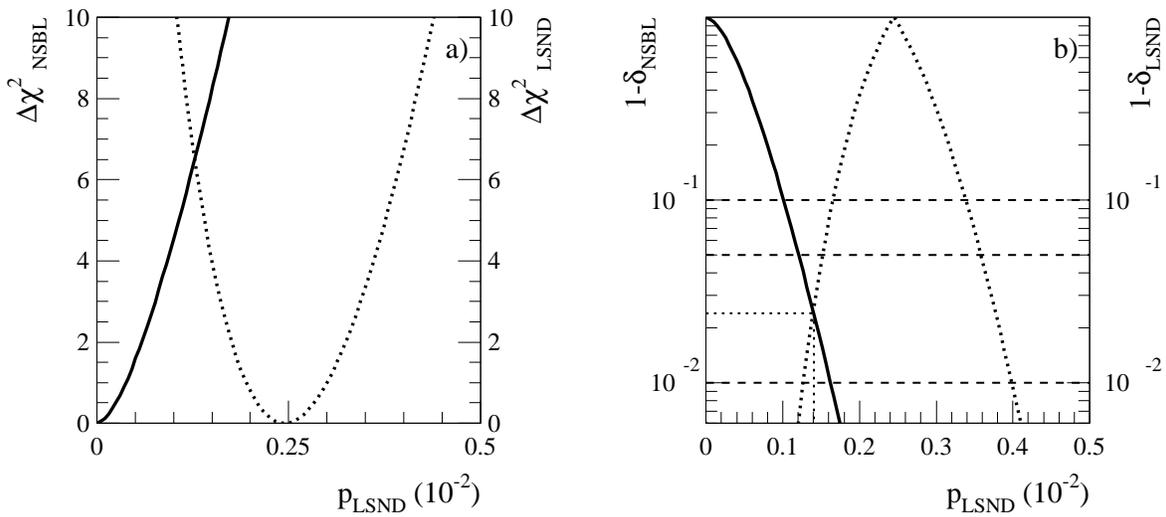}
\caption{\label{fig:fig3}a)$\chi^2$ differences
 $\Delta\chi^2_{\hbox{\tiny NSBL}}$, $\Delta\chi^2_{\hbox{\tiny LSND}}$,
 and b)
 individual confidence levels $\delta_{\hbox{\tiny NSBL}}$,
 $\delta_{\hbox{\tiny LSND}}$, as a function of the LSND
 oscillation probability $p_{\hbox{\tiny LSND}}$, for the NSBL and
 LSND datasets. The curves are for (3+1) models with the
 neutrino mass splitting $\Delta m^2$ fixed to the best-fit value
 $\Delta m^2=0.92\ \hbox{eV}^2$
 from the combined NSBL+LSND analysis, and variable mixing matrix elements
 $U_{\mu 4}$, $U_{e4}$. The solid curves refer to the NSBL dataset, the dotted ones
 to the LSND dataset.
 The dashed horizontal lines in Fig.~\ref{fig:fig3}b refer to the 90\%, 95\%,
 99\% individual confidence levels, the dotted horizontal line gives the combined
 confidence level
 $\delta =\delta_{\hbox{\tiny NSBL}}(\delta_{\hbox{\tiny LSND}}+
 (1-\delta_{\hbox{\tiny LSND}})/2)$
 for
 which the NSBL and LSND datasets are incompatible.} 
\end{figure*}
\indent We now present a slightly different approach to determine
 the statistical compatibility between the NSBL and LSND datasets in
 (3+1) models, which will prove useful in comparing the (3+1) and
 (3+2) hypotheses. \\
\indent In Fig.~\ref{fig:fig3}, we show the values for
 the $\chi^2$ differences $\Delta\chi^2_{\hbox{\tiny NSBL}}$,
 $\Delta\chi^2_{\hbox{\tiny LSND}}$, as well as the corresponding
 confidence levels $\delta_{\hbox{\tiny NSBL}}$,
 $\delta_{\hbox{\tiny LSND}}$, as a function of the LSND
 oscillation probability.
 The curves are for the set of (3+1) models with the neutrino
 mass splitting $\Delta m^2$ fixed to the best-fit value obtained in a combined
 NSBL+LSND analysis (see Section \ref{subsec:threeb}),
 $\Delta m^2=0.92\ \hbox{eV}^2$,
 and mixing matrix elements $U_{\mu 4}$, $U_{e4}$ treated as free
 parameters. The value for $\Delta m^2$ is chosen in this way
 because it represents to a good
 approximation the value for which one expects the best compatibility between
 the two datasets, as can also be seen in Fig.~\ref{fig:fig2}b.
 In Fig.~\ref{fig:fig3}a, we map the $(U_{e4},U_{\mu 4})$ allowed space
 into an the $(p_{\hbox{\tiny LSND}},\chi^2_{\hbox{\tiny NSBL}})$,
 $(p_{\hbox{\tiny LSND}},\chi^2_{\hbox{\tiny LSND}})$ spaces.
 For any given value
 of $p_{\hbox{\tiny LSND}}$,
 the minima for the $\chi^2_{\hbox{\tiny NSBL}}$ and
 $\chi^2_{\hbox{\tiny LSND}}$ functions are
 found in the two
 ($U_{e4}$, $U_{\mu 4}$) and one ($U_{e4}U_{\mu 4}$) free parameters
 available, respectively.
 The process is repeated for several ${p}_{\hbox{\tiny LSND}}$ values,
 and the collection
 of these minima for all values of $p_{\hbox{\tiny LSND}}$ give
 the two curves in Fig.~\ref{fig:fig3}a. The
 individual confidence levels $\delta_{\hbox{\tiny NSBL}}$,
 $\delta_{\hbox{\tiny LSND}}$, shown in Fig.~\ref{fig:fig3}b, are
 obtained from $\Delta\chi^2_{\hbox{\tiny NSBL}}$,
 $\Delta\chi^2_{\hbox{\tiny LSND}}$ in the usual way, by assuming
 one and two free parameters for the LSND and NSBL datasets, respectively. \\
\indent We now address how to extract areas in parameter
 space of a given combined confidence $\delta$ from
 two independent
 experimental constraints, in our case obtained via the NSBL and LSND
 datasets, without assuming statistical compatibility {\it a priori}. 
 The most straightforward way (described, for example, in
 \cite{Eitel:1999gt,Church:2002tc}) is to assign a confidence level
 $\delta =\delta_{\hbox{\tiny NSBL}}(\delta_{\hbox{\tiny LSND}}+
 (1-\delta_{\hbox{\tiny LSND}})/2)$ to the
 overlapping part (if any)
 between the two separate allowed regions in parameter space which
 are found with the constraint
 $\delta_{\hbox{\tiny NSBL}}=\delta_{\hbox{\tiny LSND}}$. The extra factor
 $(1-\delta_{\hbox{\tiny LSND}})/2$ is due to the fact that the
 LSND allowed region in the oscillation probability is two-sided, and overlap
 with the NSBL result on the same probability is obtained only for downward
 fluctuations in the LSND result, and not for upward ones. \\
\indent From Fig.~\ref{fig:fig3}b, we find overlapping allowed ranges
 in $p_{\hbox{\tiny LSND}}$ for $1-\delta_{\hbox{\tiny NSBL}}=
1-\delta_{\hbox{\tiny LSND}}\simeq 2.4\%$. We
 conclude that, in (3+1) models, the LSND and NSBL datasets are
 incompatible at a combined confidence of $\delta \simeq  
 96.4\%$. In our opinion, this value does not support
 any conclusive statements against the statistical compatibility
 between NSBL and LSND data in (3+1) models, although it represents poor
 agreement
 between the two datasets. The reader should also refer to Section
 \ref{subsec:fived}, where a different method to quantify the compatibility
 between the NSBL and LSND results is discussed.
 Future short-baseline constraints on $\nu_{\mu}\to\nu_e$ appearance,
 as well as on $\nu_{\mu}$ and $\nu_e$ disappearance, should be able to definitively establish
 whether (3+1) models are a viable solution to explain the LSND signal. 


\subsection{\label{subsec:threeb}Combined NSBL+LSND analysis}
The second analysis we perform is a combined NSBL+LSND analysis,
 with the purpose of obtaining the (3+1) allowed regions
 in parameter space, in both $(\sin^2 2\theta_{\mu e},\Delta m^2)$ and
 $(p_{\hbox{\tiny LSND}},\Delta m^2)$ space.
 A combined analysis of this sort assumes statistically compatible results.
 In Section \ref{subsec:threea}, we have shown that
 the LSND and NSBL results are marginally compatible, for (3+1) models.
 In the following, we refer to the NSBL+LSND dataset as
 the short-baseline (SBL) dataset, and we construct the
 $\chi^2$ function:
\begin{equation}
\chi^2_{\hbox{\tiny SBL}}\equiv
 \chi^2_{\hbox{\tiny NSBL}}+\chi^2_{\hbox{\tiny LSND}}
\end{equation}
\noindent where the two contributions $\chi^2_{\hbox{\tiny NSBL}}$ and
 $\chi^2_{\hbox{\tiny LSND}}$ are now simultaneously minimized
 with respect to the same set of three oscillation parameters
 $\Delta m^2$, $U_{e4}$, $U_{\mu 4}$.
\begin{figure*}[!tb]
\includegraphics*[width=18.0cm]{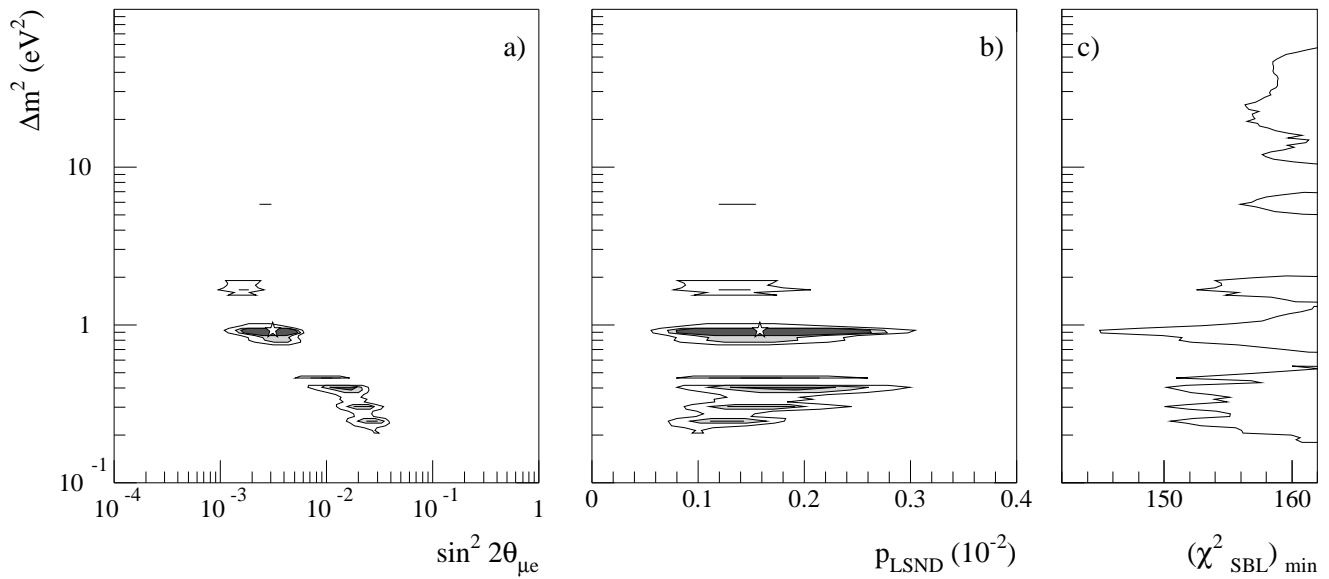}
\caption{\label{fig:fig4}Allowed regions in parameter space from a
 combined analysis of NSBL and LSND data, in (3+1) models,
 assuming statistical compatibility of the NSBL and LSND datasets.
 Fig.~\ref{fig:fig4}a shows the 90\%, 95\%, and 99\% CL allowed regions in
 $(\sin^2 2\theta_{\mu e},\Delta m^2)$ space, together with the best-fit
 point, indicated by the star; b) shows the same
 allowed regions in $(p_{\hbox{\tiny LSND}},\Delta m^2)$ space; c) shows
 the minimum $\chi^2$ value obtained in the combined analysis as
 a function of $\Delta m^2$. The number of degrees of freedom is 148.}
\end{figure*}
\begin{figure*}[!tb]
\includegraphics*[width=14.5cm]{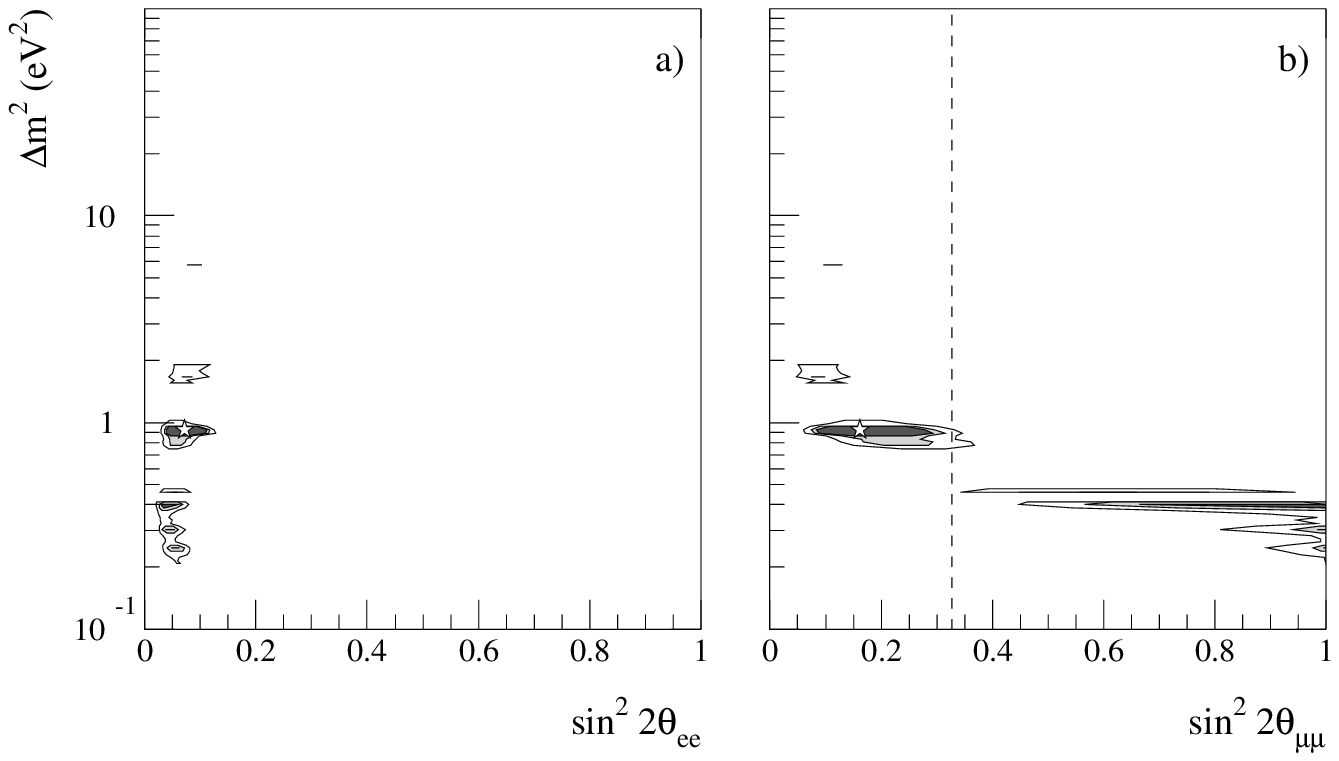}
\caption{\label{fig:fig5}Allowed regions in the parameter spaces
 relevant for $\nu_e$ and $\nu_{\mu}$ disappearance from a
 combined analysis of NSBL and LSND data, in (3+1) models,
 assuming statistical compatibility of the NSBL and LSND datasets.
 Fig.~\ref{fig:fig5}a shows the 90\%, 95\%, and 99\% CL allowed regions in
 $(\sin^2 2\theta_{e e},\Delta m^2)$ space, together with the best-fit
 point, indicated by the star; b) shows the same
 allowed regions in $(\sin^2 2\theta_{\mu\mu},\Delta m^2)$ space.
 Mixings to the right of the dashed vertical line in Fig.\ref{fig:fig5}b
 are excluded at 90\% CL by atmospheric neutrino results
 \cite{Maltoni:2001mt}, which are not included in this analysis.}
\end{figure*}
%
\indent Figs.~\ref{fig:fig4}a and \ref{fig:fig4}b show the
 90\%, 95\%, and 99\% CL 3-dimensional allowed regions in
 $(\Delta m^2,\ U_{e4},\ U_{\mu 4})$ projected onto the
 $(\sin^2 2\theta_{\mu e},\Delta m^2)$ and
 $(p_{\hbox{\tiny LSND}},\Delta m^2)$ 2-dimensional regions,
 respectively, from the combined (3+1) analysis of SBL data.
 In this combined analysis, we use the same Monte Carlo method
 described in Section \ref{subsec:threea}.
 We define the allowed regions in parameter space by performing a global
 scan. For example, the 95\% CL allowed region in the three-dimensional space
 $(\Delta m^2,\ U_{e4},\ U_{\mu 4})$ is obtained
 by requiring
$\chi^2_{\hbox{\tiny SBL}}-(\chi^2_{\hbox{\tiny SBL}})_{\hbox{\tiny min}}<
 7.82$, where
 $(\chi^2_{\hbox{\tiny SBL}})_{\hbox{\tiny min}}$ is the global
 minimum $\chi^2$ value.
 Fig.~\ref{fig:fig4}c shows the minimum $\chi^2_{\hbox{\tiny SBL}}$ values
 obtained in the combined fit, as a function of $\Delta m^2$. Of course,
 the $\chi^2_{\hbox{\tiny SBL}}$ values shown in Fig.~\ref{fig:fig4}c
 for any given $\Delta m^2$ value are larger than the sum of the
 two contributions $\chi^2_{\hbox{\tiny NSBL}}$, $\chi^2_{\hbox{\tiny LSND}}$,
 shown in Figs.~\ref{fig:fig2}c,d for the same $\Delta m^2$ value,
 since the latter were separately minimized with respect to the
 oscillation parameters. Similarly, Figs.~\ref{fig:fig5}a and \ref{fig:fig5}b
 show the projections of the 90\%, 95\%, and 99\% CL allowed regions
 in $(\Delta m^2,\ U_{e4},\ U_{\mu 4})$
 onto the $(\sin^2 2\theta_{e e},\Delta m^2)$ and
 $(\sin^2 2\theta_{\mu\mu},\Delta m^2)$ space,
 respectively, from the combined (3+1) analysis of SBL data. The zenith angle
 distribution of atmospheric muon neutrinos provides a constraint to
 $\sin^2 \theta_{\mu\mu}$ that is not included in this SBL analysis; mixings
 to the right of the dashed vertical line in
 Fig.\ref{fig:fig5}b are excluded at 90\% CL by atmospheric neutrino results
 \cite{Maltoni:2001mt}.
\indent
 The global $\chi^2$ minimum is $\chi^2_{\hbox{\tiny SBL}}$=144.9 (148
 d.o.f.). This
 $\chi^2$ value indicates an acceptable fit, assuming that the
 goodness-of-fit statistic follows the standard $\chi^2$ p.d.f. \cite{Hagiwara:fs}; for
 an alternative goodness-of-fit test, the reader should refer to Section \ref{subsec:fived}.
 The individual
 NSBL and LSND contributions to the $\chi^2$ minimum are
 $\chi^2_{\hbox{\tiny NSBL}}$=137.3 and
 $\chi^2_{\hbox{\tiny LSND}}$=7.6, respectively. This best-fit point corresponds
 to the mass and mixing parameters $\Delta m^2=0.92\ \hbox{eV}^2$, $U_{e4}=0.136$,
 $U_{\mu 4}=0.205$. 
%


\section{\label{sec:four}RESULTS FOR (3+2) MODELS}

\subsection{\label{subsec:foura}Statistical
 compatibility between NSBL and LSND}
Having introduced the relevant oscillation probability formula
 in Eq.~\ref{eq:osc_3+2}, and the statistical estimator
 $p_{\hbox{\tiny LSND}}$
 to compare the NSBL and LSND results in Section \ref{subsec:threea},
 we can now
 quantitatively address the statistical compatibility between
 the NSBL and LSND datasets under the (3+2) hypothesis. \\
\indent Ideally, we would like to determine the NSBL upper limit for
 $p_{\hbox{\tiny LSND}}$,
 for all possible combinations of the
 mass parameters $\Delta m_{41}^2$, $\Delta m_{51}^2$.
 This entails performing a scan equivalent to
 the one described in the (3+1) case as a function of $\Delta m_{41}^2$,
 shown in Fig.~\ref{fig:fig2}.
 In practice, the CPU-time requirements to pursue this route
 were prohibitive. \\
\indent An easier problem to tackle is to determine the statistical
 compatibility between the NSBL and LSND datasets only for the (3+2)
 models with mass splittings $\Delta m_{41}^2$, $\Delta m_{51}^2$
 fixed to their best-fit values, as
 obtained by the combined NSBL+LSND analysis that we present in
 Section \ref{subsec:fourb}. In sections
 \ref{subsec:threea} and \ref{subsec:threeb}, we have demonstrated that,
 at least
 for (3+1) models, this choice is a good approximation for
 the best possible statistical compatibility (see Figs.~\ref{fig:fig2} and
 \ref{fig:fig4}). \\
\begin{figure*}[!tb]
\includegraphics*[width=16.0cm]{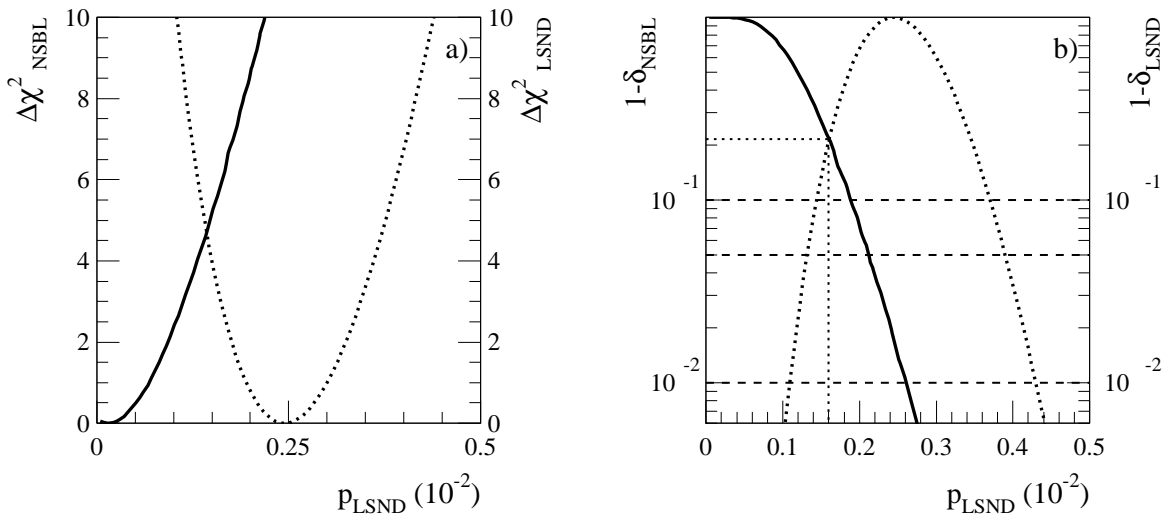}
\caption{\label{fig:fig6}a)$\chi^2$ differences
 $\Delta\chi^2_{\hbox{\tiny NSBL}}$ and $\Delta\chi^2_{\hbox{\tiny LSND}}$,
 and b) individual confidence levels $\delta_{\hbox{\tiny NSBL}}$ and
 $\delta_{\hbox{\tiny LSND}}$,
 as a function of the LSND
 oscillation probability $p_{\hbox{\tiny LSND}}$, for the NSBL and
 LSND datasets. The curves are for (3+2) models with the
 neutrino mass splittings $\Delta m_{41}^2$ and $\Delta m_{51}^2$,
 fixed to the best-fit values
 $\Delta m_{41}^2=0.92\ \hbox{eV}^2$, $\Delta m_{51}^2=22\ \hbox{eV}^2$
 from the combined NSBL+LSND analysis, and variable mixing matrix elements
 $U_{e4}$, $U_{\mu 4}$, $U_{e5}$, $U_{\mu 5}$.
 The solid curves refer to the NSBL dataset, the dotted ones
 to the LSND dataset.
 The dashed horizontal lines in Fig.~\ref{fig:fig6}b refer to the 90\%, 95\%,
 99\% individual confidence levels, the dotted horizontal line gives the combined
 confidence level
 $\delta =\delta_{\hbox{\tiny NSBL}}\delta_{\hbox{\tiny LSND}}$ for
 which the NSBL and LSND datasets are incompatible.} 
\end{figure*}
\indent In Fig.~\ref{fig:fig6}, we show the behavior of the
 $\chi^2$ values $\Delta \chi^2_{\hbox{\tiny NSBL}}$ and
 $\Delta \chi^2_{\hbox{\tiny LSND}}$, and of the confidence levels
 $\delta_{\hbox{\tiny NSBL}}$ and $\delta_{\hbox{\tiny LSND}}$, as a function
 of $p_{\hbox{\tiny LSND}}$, for the set of (3+2) models satisfying the
 requirements $\Delta m_{41}^2=0.92\ \hbox{eV}^2$,
 $\Delta m_{51}^2=22\ \hbox{eV}^2$.
 By analogy with Fig.~\ref{fig:fig3}, we map the four-dimensional
 space $(U_{e4},U_{\mu 4},U_{e5},U_{\mu 5})$ into the
 two-dimensional spaces $(p_{\hbox{\tiny LSND}},\chi^2_{\hbox{\tiny NSBL}})$
 and $(p_{\hbox{\tiny LSND}},\chi^2_{\hbox{\tiny NSBL}})$, and we plot
 the minimum $\chi^2$ values obtained for any given value of
 $p_{\hbox{\tiny LSND}}$. 
 The confidence levels shown in Fig.~\ref{fig:fig6}b are obtained from
 Fig.~\ref{fig:fig6}a considering the
 four free parameters ($U_{e4}$, $U_{\mu 4}$, $U_{e5}$, $U_{\mu 5}$)
 in the $\chi^2_{\hbox{\tiny NSBL}}$ minimization process, and the two free
 parameters ($U_{e4}U_{\mu 4}$, $U_{e5}U_{\mu 5}$) for
 $\chi^2_{\hbox{\tiny LSND}}$. \\
\indent From Fig.~\ref{fig:fig6}b, we find that, in (3+2) models, the NSBL
 and LSND datasets are incompatible at an individual confidence
 level of $\delta_{\hbox{\tiny NSBL}}=\delta_{\hbox{\tiny LSND}}=
1-0.215=78.5\%$, and at a combined confidence level
 $\delta =\delta_{\hbox{\tiny NSBL}}(\delta_{\hbox{\tiny LSND}}+
 (1-\delta_{\hbox{\tiny LSND}})/2)=70.0\%$. Fig.~\ref{fig:fig6}
 should be compared to Fig.~\ref{fig:fig3}, obtained for (3+1) models.
 A detailed comparison of the (3+1) and (3+2) hypotheses is presented
 in Section \ref{sec:five}. \\


\subsection{\label{subsec:fourb}Combined NSBL+LSND analysis}
We now turn to a combined analysis of the NSBL and LSND results in
 (3+2) models,
 assuming statistical compatibility between the two
 datasets. The purpose of this combined analysis is to obtain
 the allowed regions in the mass parameter space
 $(\Delta m_{41}^2,\Delta m_{51}^2)$, regardless of the simultaneous
 values for the mixing parameters.
 Results will be shown for $\Delta m_{51}^2\ge \Delta m_{41}^2$;
 the case $\Delta m_{41}^2>\Delta m_{51}^2$
 can be obtained by simply interchanging $\Delta m_{41}^2$ with
 $\Delta m_{51}^2$.  
 The 95\% CL allowed region is defined as the
 $(\Delta m_{41}^2,\Delta m_{51}^2)$ for which
 $\chi^2_{\hbox{\tiny SBL}}-(\chi^2_{\hbox{\tiny SBL}})_{min}< 5.99$,
 where $(\chi^2_{\hbox{\tiny SBL}})_{min}$ is the absolute $\chi^2$
 minimum for all $(\Delta m_{41}^2,\Delta m_{51}^2)$ values.
 In the minimization procedure, the mixing matrix elements $U_{e4}$,
 $U_{\mu 4}$, $U_{e5}$, $U_{\mu 5}$, are treated as free parameters.
\begin{figure}[!tb]
\includegraphics*[width=8.0cm]{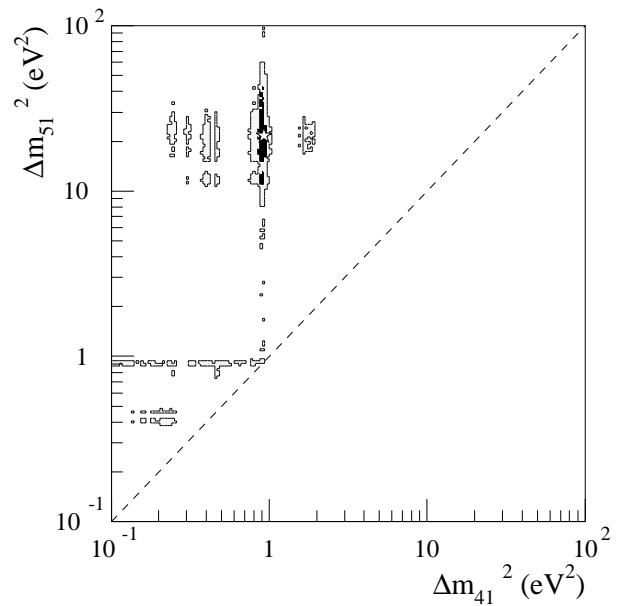}
\caption{\label{fig:fig7}Allowed ranges in
 $(\Delta m_{41}^2,\Delta m_{51}^2)$ space for (3+2) models, for
 the combined NSBL+LSND analysis, assuming statistical compatibility
 between the NSBL and LSND datasets. The star
 indicates the best-fit point, the dark and light grey-shaded regions
 indicate the 90 and 99\% CL allowed
 regions, respectively. Only the $\Delta m_{51}^2>\Delta m_{41}^2$ region is shown; the
 complementary
 region $\Delta m_{41}^2\ge\Delta m_{51}^2$ can be obtained by interchanging
 $\Delta m_{41}^2$ with $\Delta m_{51}^2$.}
\end{figure}
\indent Fig.~\ref{fig:fig7} shows the 90\% and 99\% CL allowed regions
 in $(\Delta m_{41}^2,\Delta m_{51}^2)$ space obtained in the combined
 (3+2) analysis. In light of
 the (3+1) analysis shown in previous sections, the result is not
 surprising, pointing to favored masses in the range
 $\Delta m_{41}^2\simeq 0.9\ \hbox{eV}^2$,
 $\Delta m_{51}^2\simeq 10-40\ \hbox{eV}^2$, at 90\% CL.
 At 99\% CL, the allowed region
 extends considerably, and many other
 $(\Delta m_{41}^2,\Delta m_{51}^2)$ combinations
 appear. The best-fit model ($\chi^2_{\hbox{\tiny SBL}}$=135.9, 145 d.o.f.)
 is described by the following set of parameters:
 $\Delta m_{41}^2=0.92\ \hbox{eV}^2$, $U_{e4}=0.121$, $U_{\mu 4}=0.204$,
 $\Delta m_{51}^2=22\ \hbox{eV}^2$, $U_{e5}=0.036$, and $U_{\mu 5}=0.224$.
 We note here that the best-fit is not obtained for
 fourth and fifth mass eigenstates with degenerate masses,
 that is for $\Delta m_{41}^2\simeq \Delta m_{51}^2$. The best-fit model
 we found for sub-eV neutrino masses is: $\Delta m_{41}^2=0.46\ \hbox{eV}^2$,
 $U_{e4}=0.090$, $U_{\mu 4}=0.226$, $\Delta m_{51}^2=0.89\ \hbox{eV}^2$,
 $U_{e5}=0.125$, $U_{\mu 4}=0.160$, corresponding to
 $\chi^2_{\hbox{\tiny SBL}}$=141.5 (145 d.o.f.).
%

%
%
\section{\label{sec:five} Comparing the (3+1) and (3+2) fits to SBL data}
In this section, we discuss four statistical tests that can be used
 to quantify the better overall agreement of SBL data to a
 (3+2) hypothesis for neutrino oscillations, compared to a (3+1) one.
\subsection{\label{subsec:fivea}Test 1: NSBL upper limit on
 $p_{\hbox{\tiny LSND}}$
 at a given confidence level $\delta_{\hbox{\tiny NSBL}}$}
Test 1 uses only NSBL data to establish the (3+1) and (3+2)
 upper bounds on the LSND oscillation probability
 $p_{\hbox{\tiny LSND}}$. From
 Figs.~\ref{fig:fig3} and \ref{fig:fig6}, we obtain at a confidence level
 $\delta_{\hbox{\tiny NSBL}}=0.90\ (0.99)$:
\begin{description}
\item[(3+1): ] $p_{\hbox{\tiny LSND}}<0.100\%\ (0.162\%)$
\item[(3+2): ] $p_{\hbox{\tiny LSND}}<0.186\%\ (0.262\%)$
\end{description}
Therefore, we find that (3+2) models can enhance the LSND
 probability $p_{\hbox{\tiny LSND}}$ by quite a large factor,
 compared to (3+1) models.
 The increase in $p_{\hbox{\tiny LSND}}$ that we
 obtain is significantly larger
 than the 25\% increase found in \cite{Peres:2000ic}, which is based
 on a specific choice of mixing parameters, as opposed to the complete
 parameter scan performed in this work. The value for
 the $\bar{\nu}_{\mu}\rightarrow \bar{\nu}_e$ oscillation probability
 measured by LSND \cite{lsnd} is
 $p_{\hbox{\tiny LSND}}=(0.264\pm 0.067\pm 0.045)\%$, where the errors refer
 to the 1$\sigma$ statistical and systematic errors, respectively.  

%
\subsection{\label{subsec:fiveb}Test 2: statistical compatibility between
 the NSBL and LSND datasets}
Test 2 uses both the NSBL and LSND datasets, and treats them independently to
 find the combined confidence level
 $\delta =\delta_{\hbox{\tiny NSBL}}(\delta_{\hbox{\tiny LSND}}+
 (1-\delta_{\hbox{\tiny LSND}})/2)$
 for which the datasets are incompatible, both in (3+1) and (3+2) models.
 The combined confidence levels
 can also be read from Figs.~\ref{fig:fig3} and \ref{fig:fig6}:
\begin{description}
\item[(3+1): ] $\delta = 96.4\%$
\item[(3+2): ] $\delta = 70.0\%$
\end{description}
Therefore, we find that in (3+1) models the two datasets are marginally
 compatible, and the agreement is better in (3+2) models.


\subsection{\label{subsec:fivec}Test 3: likelihood ratio test}
Test 3 combines the NSBL and LSND datasets into a single, joint analysis.
 The likelihood ratio test \cite{eadie} provides a standard way to assess
 whether two hypotheses can be distinguished in a statistically significant
 way. We define the maximum likelihood $\mathcal{L}_{i}$ from the
 minimum $\chi^2$ values $(\chi^2_{\hbox{\tiny SBL}})_{min,i}$ as
 $\mathcal{L}_{i}\equiv \exp (-(\chi^2_{\hbox{\tiny SBL}})_{min,i}/2)$,
 where the index $i=1,2$ refers to the (3+1) and (3+2) hypotheses,
 respectively. We can then form the likelihood ratio
 $\lambda_{1,2}\equiv \mathcal{L}_{1}/\mathcal{L}_{2}$. If the (3+1)
 hypothesis were as
 adequate as the (3+2) hypothesis in describing SBL data, the quantity
\begin{equation}
\chi^2_{1,2}(3)\equiv -2\ln \lambda_{1,2}=
(\chi^2_{\hbox{\tiny SBL}})_{\hbox{\tiny min,1}}
-(\chi^2_{\hbox{\tiny SBL}})_{\hbox{\tiny min,2}}
\end{equation}
\noindent should be distributed as a $\chi^2$ distribution with three
 degrees of freedom, where the number of degrees of freedom is the
 difference in the number of mass and mixing parameters in the (3+2)
 and (3+1) hypotheses, 6-3=3. \\
\indent In our combined fits, we obtain (see Sections
 \ref{subsec:threeb} and \ref{subsec:fourb}):
\begin{description}
\item[(3+1): ] $(\chi^2_{\hbox{\tiny SBL}})_{\hbox{\tiny min,1}}=144.9$,
\hspace{0.2cm} (148 d.o.f.)
\item[(3+2): ] $(\chi^2_{\hbox{\tiny SBL}})_{\hbox{\tiny min,2}}=135.9$,
\hspace{0.2cm} (145 d.o.f.)
\end{description}
\noindent and therefore $\chi^2_{1,2}(3)=9.0$. This value is significantly
 larger than 3: the probability for a $\chi^2$ distribution with three
 degrees of freedom to exceed the value $9.0$ is only 2.9\%. In other words,
 according to the likelihood ratio test, the (3+1) hypothesis should be
 rejected compared to the (3+2) one at the 97.1\% CL. Therefore,
 based on this test,
 we conclude from test 3 also that (3+2) models fit SBL data significantly
 better than (3+1) models. 
%

\subsection{\label{subsec:fived}Test 4: compatibility using the
 ``parameter goodness-of-fit''}
Test 4 uses both the results of the individual NSBL and LSND analyses, as well
 as the results of the combined NSBL+LSND analysis.
 The test is based on the ``parameter goodness-of-fit'' \cite{pgrefs} to
 compare
 the compatibility of the NSBL and LSND results under the (3+1) and (3+2)
 hypotheses. The test avoids the problem that a possible disagreement between
 the two results is diluted by data points which are insensitive
 to the mass and mixing parameters that are common to both datasets.
 The number of parameters common to both datasets is $P_c=2$ in (3+1) models,
 and $P_c=4$ in (3+2) models. One possible choice of common parameters is $(\Delta m_{41}^2, U_{e4}U_{\mu 4})$
 for (3+1) models, $(\Delta m_{41}^2, U_{e4}U_{\mu 4}, \Delta m_{51}^2, U_{e5}U_{\mu 5})$ for
 (3+2) models. The test is based on the statistic
 $\chi^2_{\hbox{\tiny PG}}= \chi^2_{\hbox{\tiny PG,NSBL}}+
 \chi^2_{\hbox{\tiny PG,LSND}}$, where
 $\chi^2_{\hbox{\tiny PG,NSBL}}\equiv
 (\chi^2_{\hbox{\tiny NSBL}})_{\hbox{\tiny SBL min}}-
(\chi^2_{\hbox{\tiny NSBL}})_{\hbox{\tiny NSBL min}}$ and
 $\chi^2_{\hbox{\tiny PG,LSND}}\equiv
 (\chi^2_{\hbox{\tiny LSND}})_{\hbox{\tiny SBL min}}-
 (\chi^2_{\hbox{\tiny LSND}})_{\hbox{\tiny LSND min}}$
 are the (positive) differences for the NSBL and LSND $\chi^2$ values
 obtained by minimizing
 the entire SBL $\chi^2$ function, minus the $\chi^2$ values
 that best fit the individual datasets.
\begin{table}[bt]
\begin{ruledtabular}
\begin{tabular}{crrrrl}
Model & $\chi^2_{\hbox{\tiny PG,NSBL}}$ &
 $\chi^2_{\hbox{\tiny PG,LSND}}$ &
 $\chi^2_{PG}$ & $P_c$ & PG (\%) \\ \hline
(3+1) & 11.8 & 4.3 & 16.1 & 2 & $3.2\cdot 10^{-2}$ \\ 
(3+2) & 7.1 & 4.4 & 11.5 & 4 & $2.1$ \\ 
\end{tabular}
\end{ruledtabular}
\caption{\label{tab:pg}Parameter goodness-of-fit PG, as defined in
 \cite{pgrefs}, to test the statistical compatibility between the NSBL
 and LSND datasets under the (3+1) and (3+2) hypotheses. The quantities
 $\chi^2_{\hbox{\tiny PG,NSBL}}$ and
 $\chi^2_{\hbox{\tiny PG,LSND}}$ are the
 NSBL and LSND contributions to the test statistic
 $\chi^2_{\hbox{\tiny PG}}$ defined in the text; $P_c$ indicates the number of
 parameters common to both datasets.}
\end{table}
\indent Table \ref{tab:pg} gives the values for the parameter goodness-of-fit
 PG as defined in \cite{pgrefs}, based on the $\chi^2_{\hbox{\tiny PG}}$
 statistic, and the number of parameters common
 to the NSBL and LSND datasets, $P_c$. This test shows a dramatic
 improvement in the compatibility between the NSBL and LSND results in going
 from a (3+1) to a (3+2) model, raising the compatibility by nearly two
 orders of
 magnitude, from 0.03\% to 2.1\%. It will be interesting to investigate
 if (3+3) models can improve the compatibility further. The resulting
 compatibility levels obtained with the parameter goodness-of-fit method are
 lower than those found in Section
 \ref{subsec:fiveb}; this, however, is not surprising,
 since the two statistical tests are quite different. 
%


\section{\label{sec:six}ADDITIONAL CONSTRAINTS}
The (3+1) and (3+2) models discussed in this work should be
 confronted with additional experimental constraints, other than the
 ones discussed in detail in the previous sections. We limit
 ourselves here to list and comment on some of these constraints,
 rather than address them in a quantitative way. Mostly, we
 will discuss the impact that such additional constraints may have
 on the best-fit (3+1) and (3+2) models found in Sections
 \ref{sec:three} and \ref{sec:four}. \\
\indent First, nonzero mixing matrix elements
 $U_{e4}$, $U_{\mu 4}$, $U_{e5}$, and $U_{\mu 5}$
 may cause observable effects in atmospheric neutrino data,
 in the form of zenith angle-independent suppressions of
 the $\nu_{\mu}$ and $\nu_e$ survival probabilities. Since
 our analysis of SBL data tends to give larger values
 for muon, rather than electron, flavor content in the fourth and fifth mass eigenstate, the effect should
 be larger on muon atmospheric neutrinos. For example,
 the (3+1) and (3+2) best-fit
 models from Sections \ref{subsec:threeb} and \ref{subsec:fourb}
 would give an overall suppression of the $\nu_{\mu}$ flux of
 8\% and 17\%, respectively. The size of the effect of
 $\nu_{\mu}\rightarrow \nu_x$ oscillations at high $\Delta m^2$
 is comparable
 to the current accuracy with which the absolute normalization
 of the atmospheric neutrino flux is known \cite{atmflux},
 which is approximately 20\%. A more quantitative analysis using the full
 Super-Kamiokande and MACRO spectral information \cite{Maltoni:2001mt}
 puts an upper bound of 16\% at 90\% CL on this high $\Delta m^2$
 contribution to the atmospheric $\nu_{\mu}$ flux suppression (in the
 notation of Ref. \cite{Maltoni:2001mt}, this suppression is parametrized
 as $2d_{\mu}(1-d_{\mu})$, where $d_{\mu}<0.09$ at 90\% CL).
 Therefore, it is expected
 that the inclusion of atmospheric neutrino data in this analysis would pull
 the best-fit muon flavor components in the fourth and fifth mass eigenstates
 to lower values, but not in a dramatic way (see also Fig. \ref{fig:fig5}b). \\
\indent Second, models with large masses $m_4$ and $m_5$, and with nonzero
 mixing matrix elements $U_{e4}$ and $U_{e5}$,
 should be confronted with tritium $\beta$ decay measurements.
 The presence of neutrino
 masses $m_4$ and $m_5$ introduces kinks in the differential $\beta$
 spectrum; the location in energy of the kinks is determined by the neutrino
 masses, and the size of the kinks is determined by the amount of electron
 flavor component in the fourth and fifth mass eigenstates. 
 For a spectrometer
 integrating over the electron energy interval $\delta$ near the
 $\beta$-decay endpoint, the count rate is \cite{tritium_phenomenology}:
\begin{equation}
n(\delta ) = \frac{\bar{R}}{3}\sum_{i=1}^n U_{ei}^2(\delta^2-m_i^2)^{3/2}
\label{eq:tritium_exact}
\end{equation}
\noindent where the quantity $\bar{R}$ does not depend on the small neutrino
 masses and mixings, $n=4$ or $n=5$ for (3+1) or (3+2) models, respectively,
 and we have assumed $\delta > m_i,\ i=1,\ldots ,n$, and
 CP invariance. From the experimental point of view, tritium $\beta$ decay
 results are generally expressed in terms of a single effective mass
 $m(\nu_e)$:
\begin{equation}
n_s(\delta ) = \frac{\bar{R}}{3} (\delta^2-m(\nu_e)^2)^{3/2}
\label{eq:tritium_onemass}
\end{equation}
\noindent where $m(\nu_e)$ is the fit mass parameter. In the limit $\delta^2
 \gg m_i^2,\ i=1,\ldots ,n$ the relation between the true masses and
 mixings to the fitted mass $m(\nu_e)$ is independent from the integration
 interval $\delta$:
\begin{equation}
m(\nu_e)^2\simeq \sum_{i=1}^n U_{ei}^2m_i^2
\label{eq:tritium_largedeltalimit}
\end{equation}
The condition $\delta^2 \gg m_i^2,\ i=1,\ldots ,n$ is generally satisfied for
 the neutrino masses considered in this paper, in order to ensure sufficient
 $\beta$ decay count rate statistics in the experiments.
 Therefore, to a first approximation, we can consider the effect
 of heavy neutrino masses $m_4$, $m_5$ only on the single mass parameter
 $m(\nu_e)$ fitted by the experiments. A more general analysis assessing
 the sensitivity of current and future $\beta$ decay experiments to
 multiple fitted neutrino masses, although highly
 desirable, is beyond the scope of this work; for further details, the reader
 should consult Ref. \cite{tritium_phenomenology}.
The current best measurements on $m(\nu_e)^2$ come from the Troitsk and Mainz
 experiments \cite{current_tritium_results}, which have very similar
 $m(\nu_e)^2$ sensitivities. Both found no evidence
 for a nonzero $m(\nu_e)^2$ value; the latest Mainz result is
 $m(\nu_e)^2=-1.6\pm 2.5\pm 2.1\ \hbox{eV}^2$, or $m(\nu_e)\le
 2.2$ eV at 95\% CL, using $\delta =70$ eV \cite{current_tritium_results}.
 Now, assuming a normal hierarchy ($m_1<m_4<m_5$) with $m_1\simeq 0$, the
 $\beta$ decay neutrino mass in Eq. \ref{eq:tritium_largedeltalimit} can be
 written as $m(\nu_e)\simeq U_{e4}^2\Delta m^2_{41}+U_{e5}^2 \Delta m^2_{51}$;
 the best-fit (3+1) and
 (3+2) models found in this analysis would give $m(\nu_e)^2=0.017\ \hbox{eV}^2$
 and $m(\nu_e)^2=0.042\ \hbox{eV}^2$, respectively, that is $m(\nu_e)^2$ values
 well below the current experimental sensitivity. The planned tritium
 $\beta$ decay experiment KATRIN should be able to improve the sensitivity to
 $m(\nu_e)$ by roughly an order of magnitude in the forthcoming
 years, thanks to its better statistics, energy resolution, and
 background rejection \cite{tritium_katrin}. Specifically, the
 systematic and statistical (for $\delta \gtrsim 30\ \hbox{eV}$)
 uncertainties on the single fitted mass $m(\nu_e)^2$ quoted for KATRIN are
 0.018 and 0.016 eV$^2$, respectively \cite{tritium_katrin}, which should
 provide some sensitivity to the preferred (3+1) and
 (3+2) neutrino models with a normal mass hierarchy,
 $m_1<m_4<m_5$.
 We now consider mass spectra with an inverted hierarchy, defined here as
 $m_4<m_1$ for (3+1) models, and $m_5<m_4<m_1$ for (3+2) models. We note
 that for (3+2) models other hierarchies are also possible, but those
 do not satisfy the implicit assumption $|\Delta m^2_{51}|=
 |\Delta m^2_{54}|+|\Delta m^2_{41}|$ taken in this analysis.
 The $\beta$ decay neutrino mass in Eq.
 \ref{eq:tritium_largedeltalimit} can now be written as $m(\nu_e)^2\simeq
 |\Delta m^2_{41}|$ for inverted (3+1) models, and as
 $m(\nu_e)^2\simeq |\Delta m^2_{51}|$ for inverted (3+2) models.
 Clearly, in this case the $\beta$
 decay constraints depend strongly on the values of
 $|\Delta m^2_{41}|,\ |\Delta m^2_{51}|$, and models with $\gtrsim 5\
 \hbox{eV}^2$ mass splittings are already severely disfavored. \\
\indent Third, introducing sterile neutrinos may affect a number
 of cosmological predictions, which are derived from various
 measurements \cite{cosmoobs}. The standard cosmological
 model predicts that
 sterile neutrinos in the $\sim 1\ \hbox{eV}$ mass range with a significant
 mixing with active neutrinos would be present
 in the early Universe with the same abundance as the active
 neutrino species, in disagreement
 with cosmological observations \cite{Abazajian:2002bj,DiBari}.
 On the other hand,
 several models have been
 proposed that would reconcile sterile neutrinos with cosmological
 observations, for example suppressing
 thermal abundances for sterile neutrinos
 (see, {\it e.g.}, Ref.\cite{Abazajian:2002bj} and references
 therein). In particular, active-sterile oscillations in
 the early Universe
 may provide a natural mechanism to suppress the relic abundances of
 sterile neutrinos \cite{Foot:1995qk}, and scenarios invoking multiple
 sterile neutrinos are being investigated \cite{DiBari}.


\section{\label{sec:seven}CONCLUSIONS}
We have performed a combined analysis of seven
 short-baseline experiments (Bugey, CHOOZ, CCFR84, CDHS, KARMEN,
 LSND, NOMAD) for both the (3+1) and the (3+2) neutrino oscillation hypotheses,
 involving one and two sterile neutrinos at high $\Delta m^2$,
 respectively. The motivation for considering more than one sterile
 neutrino arises from the tension in trying to reconcile,
 in a CPT-conserving, four-neutrino picture, the LSND signal for oscillations
 with the null results obtained by the other short-baseline experiments.
 Multiple ({\it e.g.} three) sterile neutrinos can also be motivated on
 theoretical grounds. \\
\indent We have described two types of analyses for both
 the (3+1) and (3+2) neutrino oscillation hypotheses. In the first analysis,
 we treat the LSND and the null short-baseline (NSBL) datasets separately,
 and we determine the statistical compatibility between the two. In
 the second analysis, we assume statistical compatibility and
 we combine the two datasets, to obtain the
 favored regions in neutrino mass and mixing parameter space. \\
\indent The main results of the analysis are summarized in Section
 \ref{sec:five}, where we compare the adequacy of the (3+1) and (3+2)
 hypotheses in describing neutrino short-baseline data,
 by means of four statistical tests.  
 First, we treat the LSND oscillation probability as
 a parameter that can be measured with NSBL data alone, and find that
 the NSBL 90\% CL upper limit on the LSND oscillation probability  can be
 significantly relaxed by going from (3+1) to (3+2) models,
 by about 80\%. Second,
 the combined confidence level for which the NSBL and LSND datasets
 are incompatible is determined to be 96.4\% and 70.0\% in the analysis, 
 for the (3+1) and (3+2)
 hypotheses, respectively. Third, a likelihood ratio test of the
 two hypotheses is discussed, and shows that the (3+1) hypothesis should be
 rejected compared to the (3+2) one at the 97.1\% CL. Fourth,
 the parameter ``goodness-of-fit'' defined in \cite{pgrefs} shows much
 better agreement between the NSBL and LSND results for (3+2)
 models than for (3+1) models. \\
\indent In conclusion, we find that (3+1) models are only marginally allowed
 when considering all of the seven short-baseline results, including LSND,
 in agreement with previous analyses
 \cite{Peres:2000ic,Strumia:2002fw,Grimus:2001mn},
 and that (3+2) models can provide a better description of the data.
 Only the simplest neutrino mass and mixing patterns have been fully
 characterized in the literature so far, and the analysis described in this
 paper may be viewed as a simple attempt
 to explore more generic scenarios, which appear both experimentally
 and theoretically plausible.
 Given the bright potential for precision measurements by neutrino
 oscillation experiments in the near future, a more general
 phenomenological approach may be needed.
 


\begin{acknowledgments}
 We thank J.~Bouchez, L.~Camilleri, K.~Eitel, J.J.~Gomez-Cadenas,
 E.~A.~Hawker, G.~B.~Mills, E.~Nagy,
 V.~Valuev, and G.~P.~Zeller for kindly providing data used in this analysis.
 We thank K.~N.~Abazajian, G.~Barenboim,
 S.~J.~Brice, K.~Eitel, B.~Kayser, W.~C.~Louis, M.~Maltoni, J.~Monroe,
 P.~Nienaber, T.~Schwetz, A.~Y.~Smirnov, and K.~Whisnant for valuable
 discussions and useful suggestions. This work was supported by NSF.
\end{acknowledgments}

\appendix
\section{\label{sec:eight}PHYSICS AND STATISTICAL ASSUMPTIONS}

\indent In this section, we briefly describe the physics and statistical
 assumptions used to obtain the approximate characterizations of the
 short-baseline experiments used in the analysis.
 For the analysis of the
 Bugey, CDHS, and KARMEN data, we also refer to the excellent reference
 \cite{Grimus:2001mn}, which we followed closely.


%
\begin{figure*}[!tb]
\includegraphics[width=8.0cm]{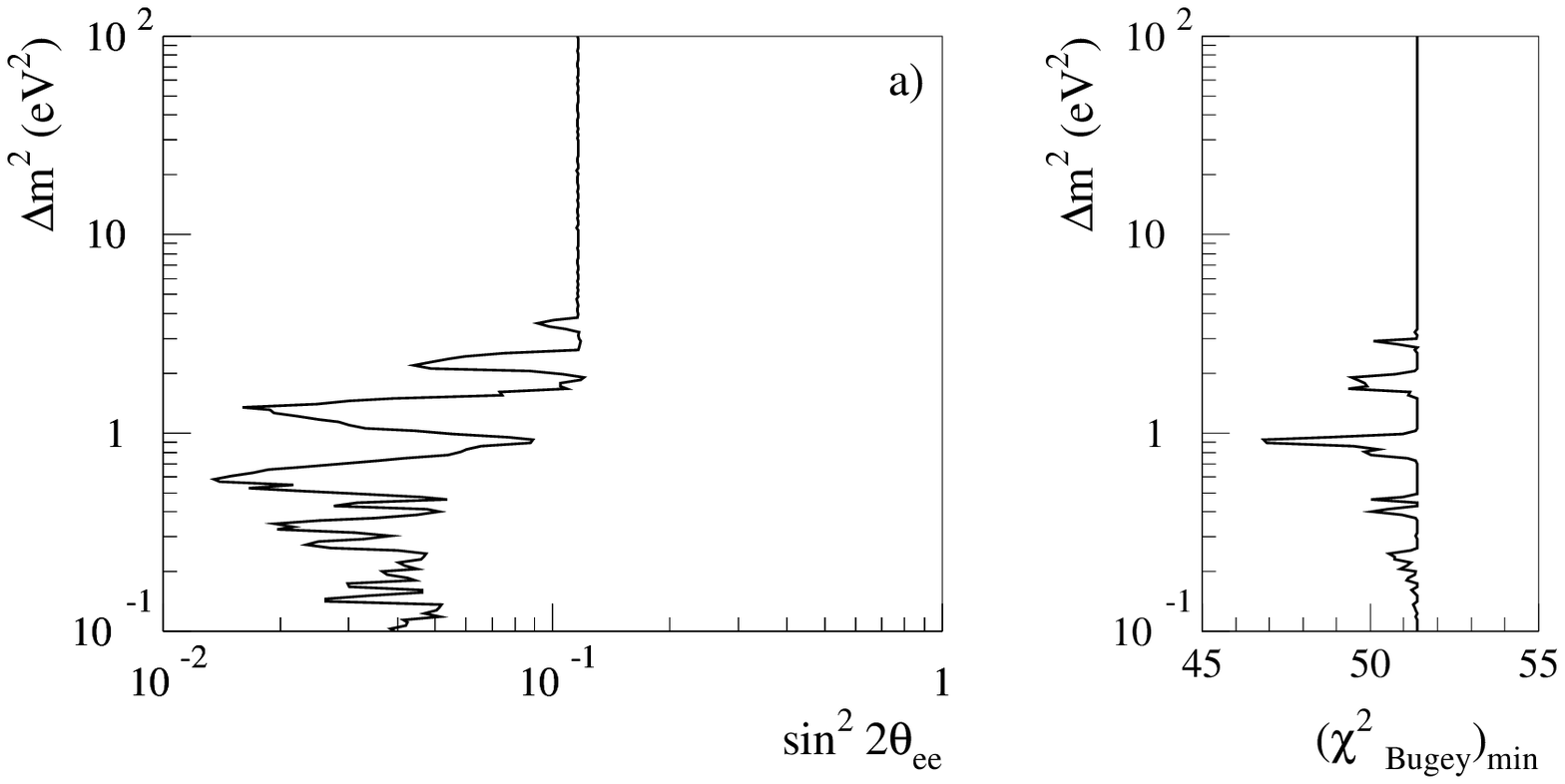} \hspace{1.5cm}
\includegraphics[width=8.0cm]{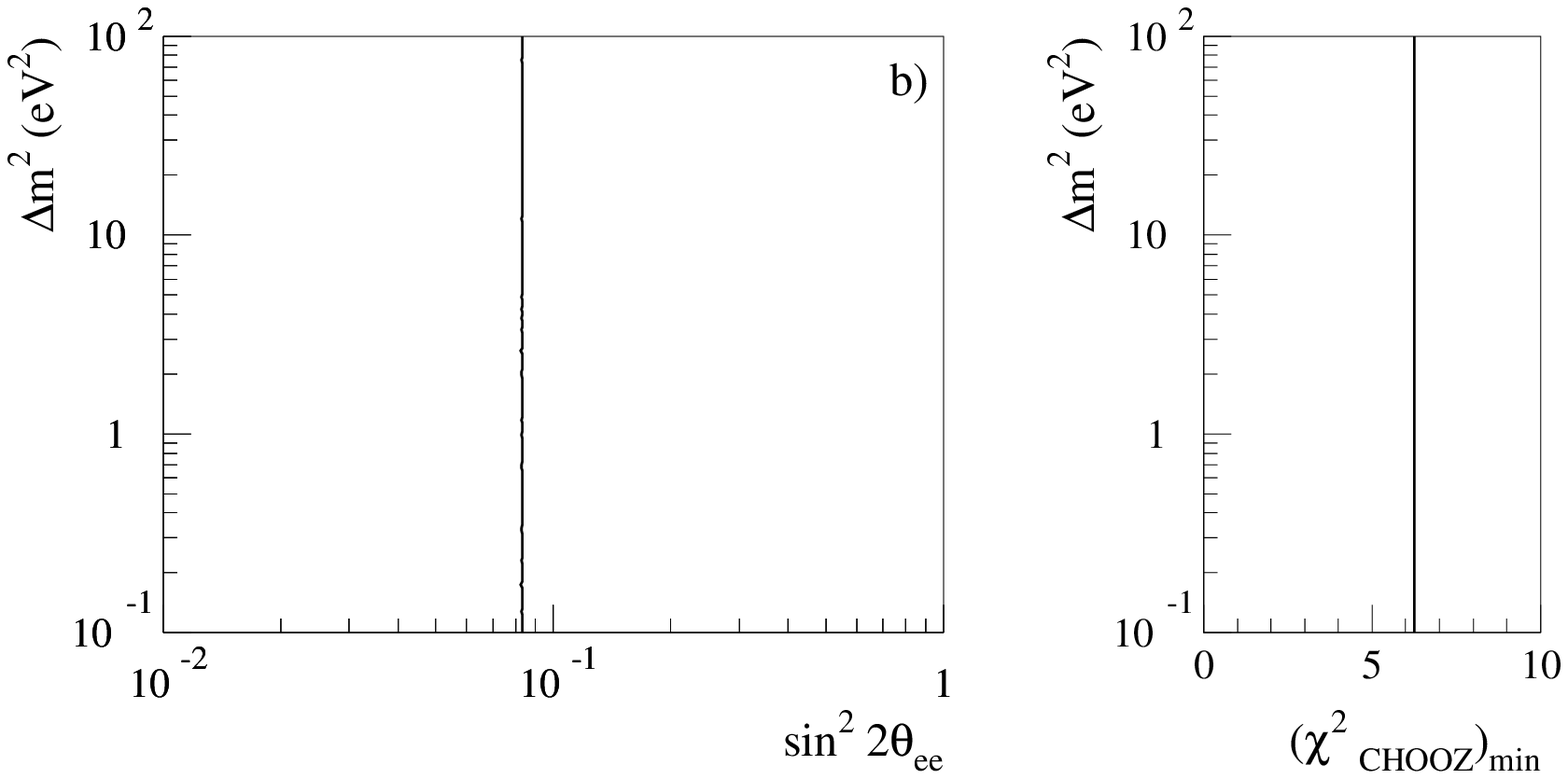}
\includegraphics[width=8.0cm]{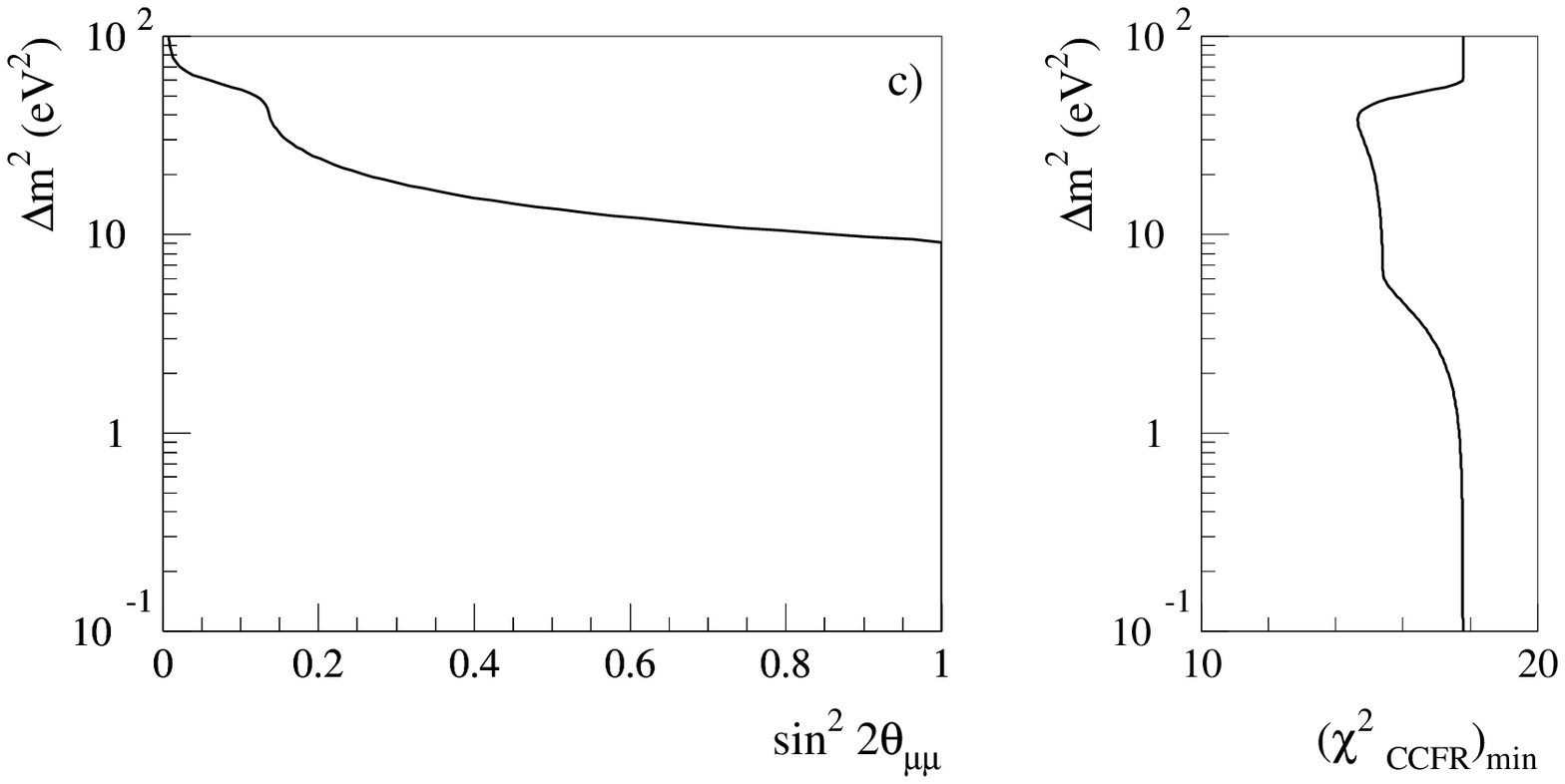} \hspace{1.5cm}
\includegraphics[width=8.0cm]{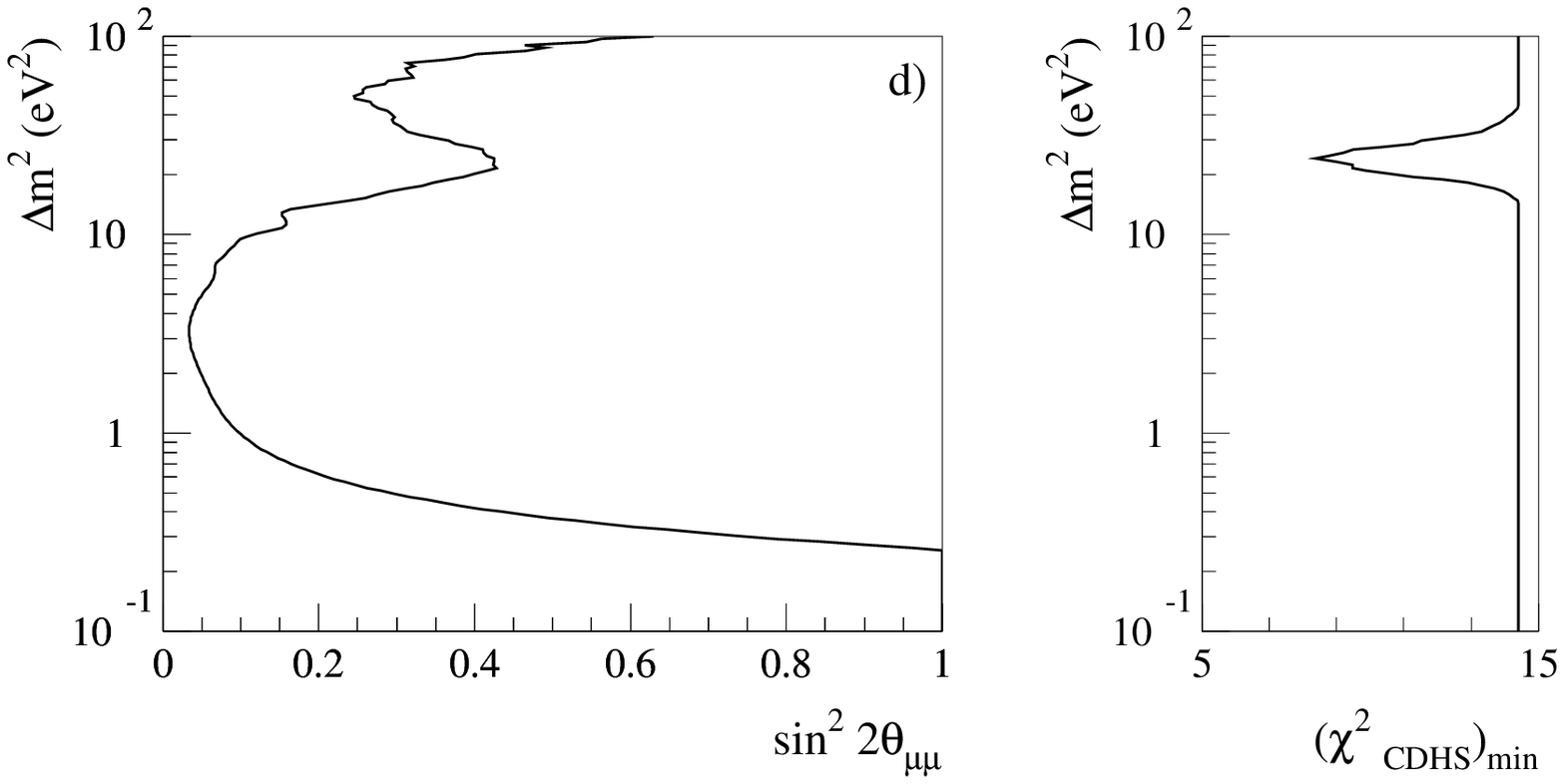}
\includegraphics[width=8.0cm]{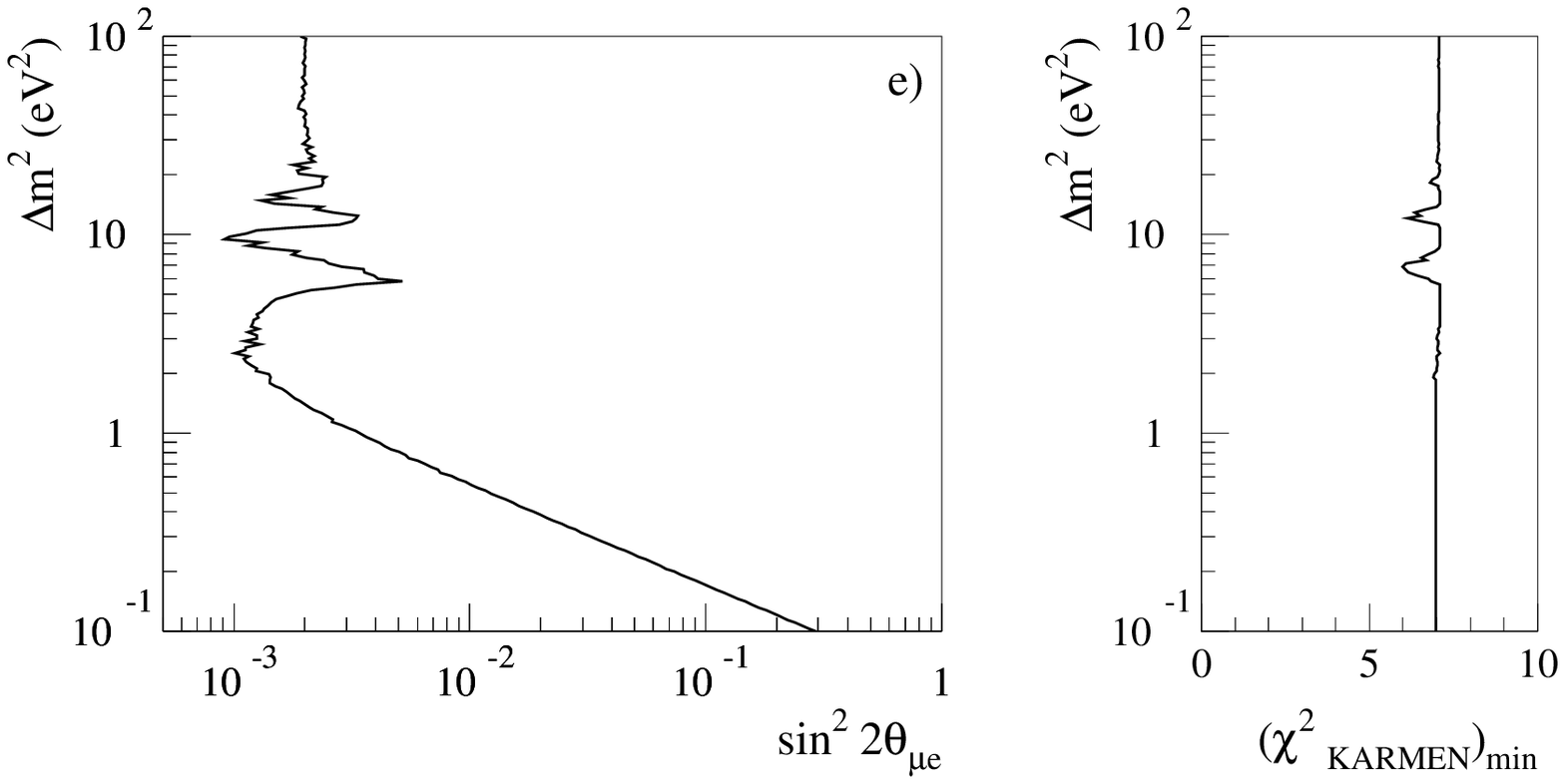} \hspace{1.5cm}
\includegraphics[width=8.0cm]{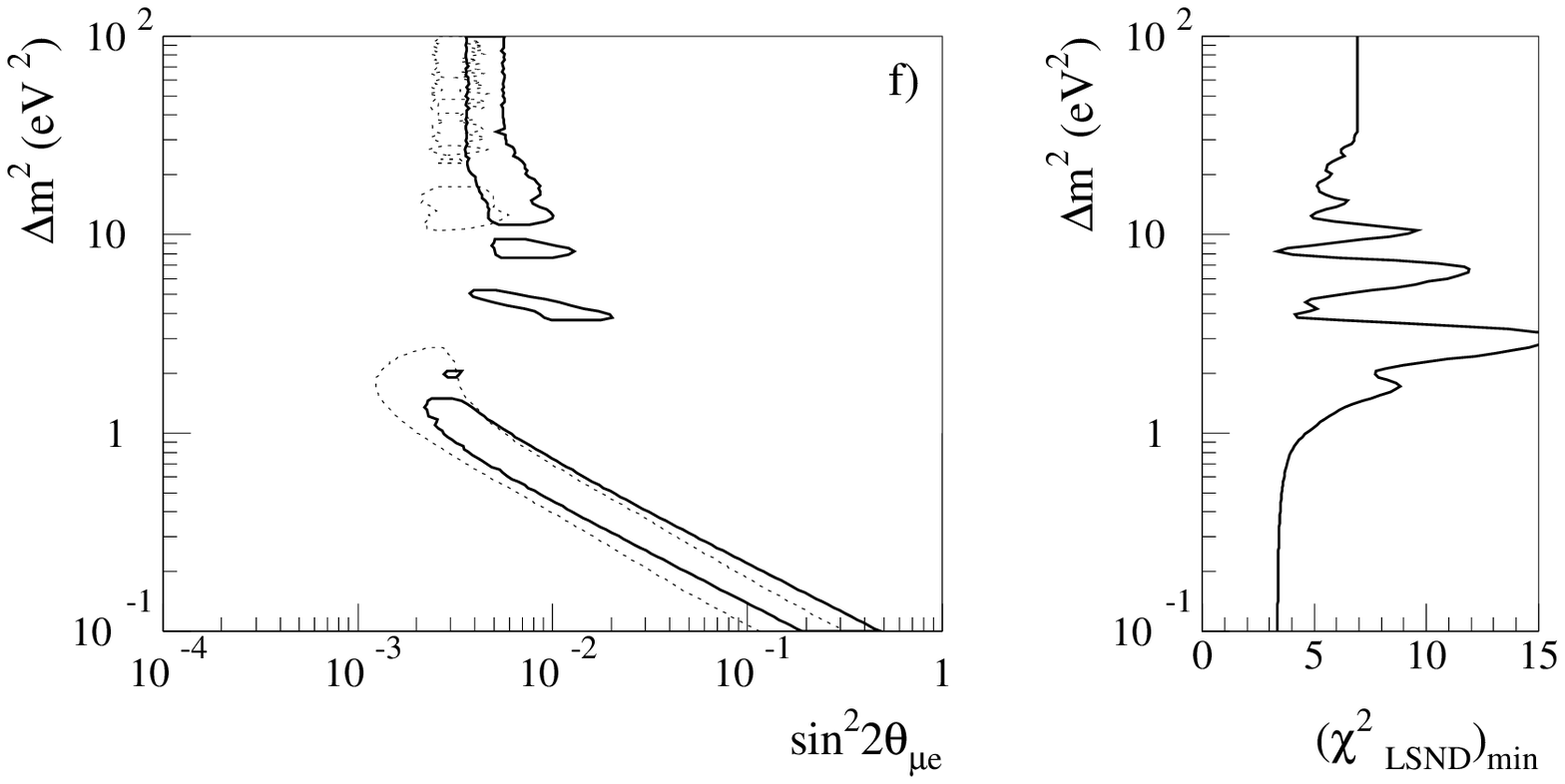}
\includegraphics[width=8.0cm]{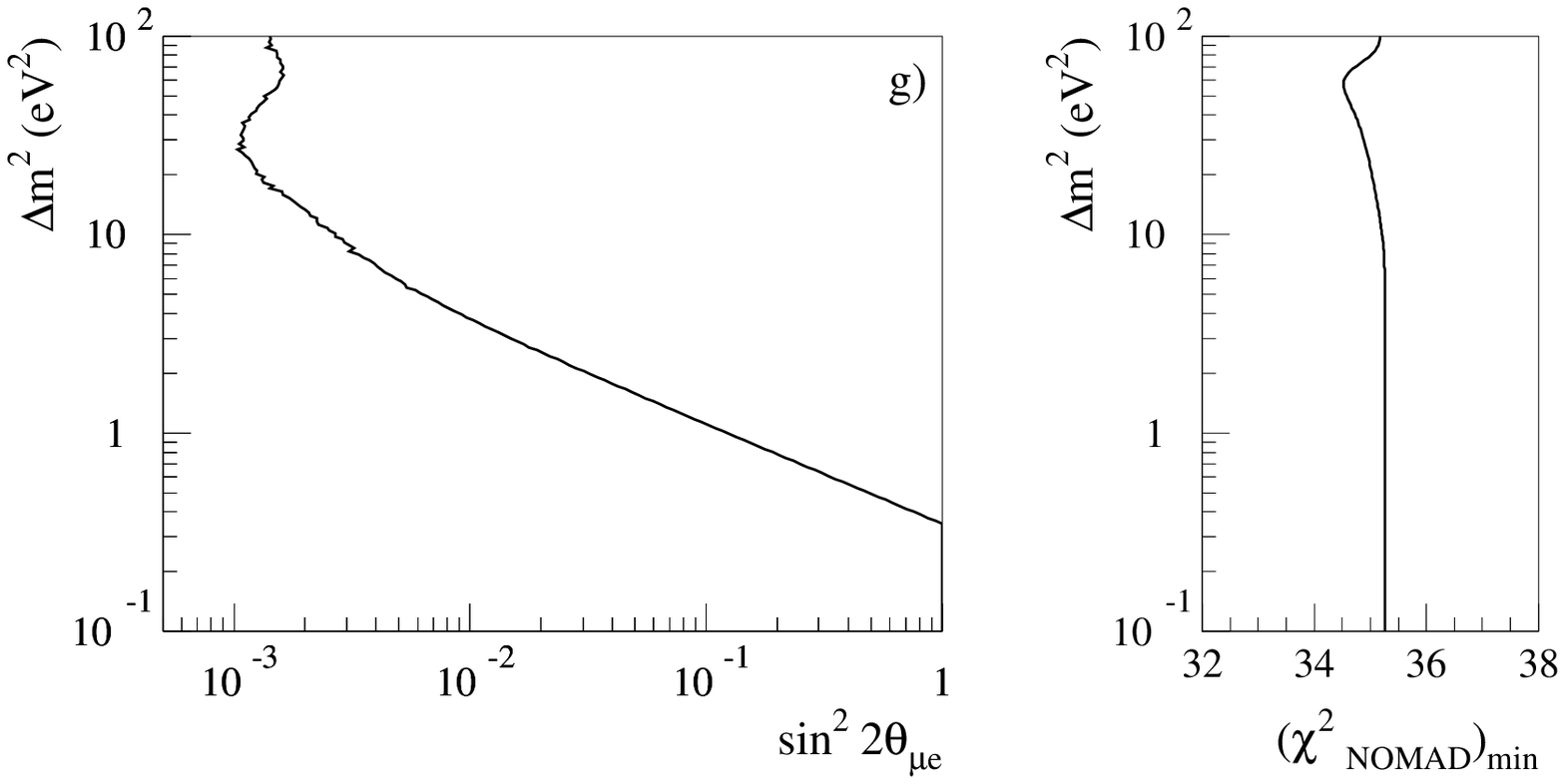} \hspace{1.5cm}
\makebox[8.cm][l]{}
\caption{\label{fig:fig8}90\% CL upper limits on oscillations derived
 in this analysis for the following NSBL experiments: a) Bugey,
 b) CHOOZ,
 c) CCFR84, d) CDHS, e) KARMEN, g) NOMAD.
 Fig.~\ref{fig:fig8}f shows the LSND
 90\% CL allowed region obtained with the decay-at-rest analysis
 described in Appendix \ref{sec:eight} (solid line),
 superimposed to the published LSND 90\% CL allowed region
 (dashed line).
 Also shown are
 the $(\chi^2)_{\hbox{\tiny min}}$ values as a function of $\Delta m^2$
 obtained by all the experiments considered individually.
 The number of degrees of freedom is 58 in Bugey, 12 in CHOOZ, 16 in CCFR84,
 13 in CDHS, 7 in KARMEN, 3 in LSND, 28 in NOMAD.} 
\end{figure*}
\indent The Bugey experiment \cite{Declais:1994su} is sensitive to
 $\bar{\nu}_e$ disappearance by
 measuring the charged-current interaction
 of $\bar{\nu}_e$'s produced by two nuclear reactors at the Bugey nuclear
 power plant. Two liquid scintillator
 detectors, located at different positions, are used.
 The signature for an antineutrino interaction is a positron and a
 delayed light pulse produced by the subsequent neutron capture on
 $^6\hbox{Li}$.
 Data are given for three baselines: $15$, $40$, and $95$ m between
 neutrino production and detection. We follow the ``normalized
 energy spectra'' analysis discussed in the Bugey paper \cite{Declais:1994su}. 
 The data are presented as ratios of observed to
 predicted (for no oscillations) positron energy spectra, between $1$ and
 $6$ MeV positron energy. We use 25,25, and 10 positron energy bins
 for the $15$, $40$, $95$ m baselines, respectively.
 In the $\chi^2$ analysis,
 fits included not only the mass and mixing parameters, but also five large
 scale deformations of the positron spectrum due to
 systematic effects. The experimental positron energy resolution and
 the neutrino baseline smearing are taken into account; the neutrino
 cross-section energy dependence within a positron energy bin
 is not (the energy bin widths are small). \\
\indent Similarly, the CHOOZ experiment \cite{Apollonio:2002gd}
 investigates $\bar{\nu}_e$
 disappearance by observing interactions of
 $\bar{\nu}_e$'s produced by two nuclear reactors $\simeq 1$ km away
 from the CHOOZ detector. The signature for a neutrino
 interaction is a delayed coincidence between the prompt $e^+$ signal
 and the signal due to the neutron capture in the Gd-loaded scintillator.
 We follow ``analysis A,'' as discussed in the CHOOZ paper
 \cite{Apollonio:2002gd}. Data
 are given as positron yields as a function of
 energy. In this analysis, seven positron energy bins, between
 $0.8$ and $6.4$ MeV, are considered, for which the CHOOZ observations, as well
 as the
 predictions on the positron yields for the no-oscillation case from both
 reactors, are given in \cite{Apollonio:2002gd}.
 Because of the presence of two reactor sources, the $\chi^2$ analysis
 comprises 14 positron
 yield bins for a given energy/baseline. We use the full covariance matrix
 to take into account the fact that the yields corresponding to the same
 energy bin are extracted for both reactors simultaneously, as is done in
 \cite{Apollonio:2002gd}. The analysis fits for the systematic uncertainty
 in the absolute normalization constant on the $\bar{\nu}_e$ yield
 from the reactors, in addition to the mass and mixing parameters.
 Since we are interested in the $\Delta m^2>0.1\ \hbox{eV}^2$
 range only, where no energy shape distortions are expected, we neglect
 the systematic uncertainty on the energy-scale calibration, and the effect
 of the positron energy resolution. \\
\indent The CCFR84 experiment \cite{Stockdale:1984cg} constrains
 $\nu_{\mu}$ and $\bar{\nu}_{\mu}$
 disappearance by measuring the charged-current interaction
 of muon neutrinos and antineutrinos, produced by a Fermilab secondary,
 sign selected beam yielding
 $40<E_{\nu}<230$ GeV neutrinos from $\pi^{\pm}$ and $K^{\pm}$ decays
 in the $352$ m long decay pipe. We refer here to the 1984 CCFR experiment
 (hence the label CCFR84 throughout the text),
 which operated with two similar detectors located at different
 distances from the neutrino source, $715$ and $1116$ m from the
 mid-point of the decay region, respectively. The two sampling calorimeter
 detectors
 consisted of steel plates and scintillation counters.
 Six secondary beam momentum settings were used: five for neutrino running,
 and one for antineutrino running. For each secondary beam momentum setting,
 the data are divided into
 three neutrino energy bins, for a total of eighteen energy bins,
 from Ref.\cite{Stockdale:1984ap}. Data
 are presented as double ratios: the
 far to near detector ratio of observed number of events, divided by the
 far to near ratio of events predicted for no oscillations. As in
 \cite{Stockdale:1984cg}, only the mean neutrino energy for a given
 neutrino energy bin
 is used in the $\chi^2$ analysis. The systematic and statistical
 uncertainties on the
 far to near ratio normalization are taken into account. The systematic
 uncertainty is assumed
 to be energy-independent and totally correlated between any two energy bins.
 The neutrino pathlength smearing, mostly due
 to the long decay region, is also taken into account. \\ 
\indent The CDHS experiment \cite{Dydak:1983zq} is also sensitive to
 $\nu_{\mu}$ disappearance
 via the charged-current interaction
 of $\nu_{\mu}$'s, produced by a 19.2 GeV/c proton beam from the CERN
 Proton Synchrotron. Two detectors are located at $130$ and $835$ m from
 the target. The detectors are sampling calorimeters, with
 iron and scintillator modules interspersed, to measure
 the range of a muon produced in a neutrino interaction. Fifteen muon range
 bins are used. The data are presented as double ratios: the
 far to near detector ratio of the observed number of events, divided by the
 far to near ratio of the number of events predicted for no oscillations.
 Neutrino
 energy distributions are obtained for a given muon energy (or range) via the
 NUANCE \cite{Casper:2002sd} neutrino cross-section generator. As for CCFR84,
 the
 systematic
 uncertainty on the far to near ratio and the neutrino baseline smearing are
 taken into account. \\
\indent The KARMEN experiment \cite{Armbruster:2002mp} investigates the
 $\bar{\nu}_{\mu}\rightarrow \bar{\nu}_e$ appearance channel, from
 $\bar{\nu}_{\mu}$'s produced in the $\pi^+$-$\mu^+$-decay at
 rest (DAR) chain of the ISIS neutrino source. KARMEN 
 measures the charged-current interaction
 $p(\bar{\nu}_e,e^+)n$, with a liquid scinitillator detector located
 at an average distance $17.7$ m downstream of the neutrino source.
 The $\bar{\nu}_e$ signature is a spatially correlated delayed
 coincidence between a prompt positron and a delayed $\gamma$
 event from a $(n,\gamma )$ neutron capture reaction. In this
 analysis, only the positron (``prompt'') energy distribution
 after all cuts is taken into account, given in \cite{Armbruster:2002mp}.
 The data are binned into nine prompt energy bins, between $16$ and
 $50$ MeV (all bins are $4$ MeV wide, except the highest energy
 one, ranging from $48$ to $50$ MeV). In predicting the prompt
 energy distribution for a set of mass and mixing oscillation parameters,
 the given Monte Carlo positron energy distribution, and the total number of
 events expected after all cuts for full mixing and $\Delta m^2=100\ \hbox{eV}^2$,
 are used \cite{Eitel:1998iv}. Energy resolution and baseline smearing effects
 (due to finite detector size) are taken into account. Given the
 low statistics of the nine KARMEN prompt energy bins, we construct
 the $\chi^2$ function by first defining the
 likelihood ratio \cite{Hagiwara:fs}:
\begin{equation}
\lambda ({\bf \theta} )=\frac{f({\bf n}; {\bf \mu}({\bf \theta}),
 {\bf b})}{f({\bf n}; {\bf n}, {\bf b})}
\end{equation}
\noindent where ${\bf \theta}$ denotes schematically all mass and mixing
 parameters, ${\bf n}$, ${\bf \mu}({\bf \theta})$ and ${\bf b}$ are
 the data, expected signal, and expected background vectors with
 nine elements, and $f({\bf n}; {\bf \mu}({\bf\theta}), {\bf b})$
 are the probabilities for a Poisson process with known background:
\begin{equation}
f({\bf n}; {\bf\mu}({\bf\theta}), {\bf b})=
\prod_{i=1}^9 \frac{(\mu_i+b_i)^{n_i}\exp(-(\mu_i+b_i))}{n_i!}
\end{equation} 
We define $\chi^2_{\hbox{\tiny KARMEN}}$ as:
\begin{equation}
\chi^2_{\hbox{\tiny KARMEN}}\equiv -2\ln \lambda ({\bf\theta})
\end{equation}
%
\indent The LSND experiment at Los Alamos \cite{lsnd} is also sensitive to
 $\bar{\nu}_{\mu}\rightarrow \bar{\nu}_e$ appearance, with a
 neutrino source and detection signature similar to that of KARMEN, but with
 better statistics. The LSND
 liquid scintillator detector is located at an average distance of $30$ m
 from the neutrino source. As for KARMEN, in this analysis we consider
 only the positron energy distribution arising from a $\bar{\nu}_e$
 interaction in mineral oil, published as five energy bins between
 $20$ and $60$ MeV \cite{lsnd}. Our analysis ignores the
 information arising from the higher-energy neutrinos from pions decaying
 in flight, which has a smaller (but non-negligible) sensitivity to
 oscillations compared to the decay at rest (DAR) sample considered here.
 In our simulation, we take into account the expected
 energy distribution from $\mu^+$ decay at rest, the neutrino baseline
 distribution 
 for the $8$ m long cylindrical detector, the neutrino energy dependence 
 of the cross-section for the detection process $p(\bar{\nu}_e,e^+)n$
 (including nuclear effects, simulated with the NUANCE \cite{Casper:2002sd}
 neutrino cross-section
 generator), and the experimental energy resolution.
 We use the published numbers for the background expectations, the
 number of $\bar{\nu}_e$ events for 100\%
 $\bar{\nu}_{\mu}\rightarrow \bar{\nu}_e$ transmutation, and
 for the efficiency of the event selection criteria. We construct the LSND
 $\chi^2$ function in the same way as we construct the one for KARMEN, because
 of the low statistics of the data sample. \\
\indent Finally, the NOMAD experiment is sensitive to
 $\nu_{\mu}\to\nu_e$ oscillations at $\Delta m^2\gtrsim 1\ \hbox{eV}^2$ by
 looking for charged-current muon neutrino and electron neutrino
 interactions in the NOMAD detector \cite{nomad}. The detector
 consists of a
 large dipole magnet which houses drift chambers to measure the momenta of the
 charged particles produced in neutrino interactions; transition radiation
 modules for lepton identification; an electromagnetic calorimeter to
 measure the energy of
 electrons and photons; a hadron calorimeter for particle
 identification;
 and muon chambers for muon identification. Neutrinos are produced by
 impinging 450 GeV protons extracted from the CERN SPS accelerator onto
 a thick beryllium target. The secondary particles produced in the target
 are focused into a nearly parallel beam by two magnetic lenses, and decay
 in a 290 m long decay tunnel to produce a $\sim 10-100$ GeV neutrino beam
 with about 1\% $\nu_e$ contamination. Neutrino interactions are then
 observed in the NOMAD detector at an average distance of 625 m from the
 neutrino source. The $\nu_{\mu}\to\nu_e$ search is
 performed by comparing the measured ratio $R_{e\mu}$ of the number of $\nu_e$
 to $\nu_{\mu}$ charged-current neutrino interactions with the one expected
 in the absence of oscillations. The data are binned into 30 bins,
 covering ten bins in visible energy between 3 and 170 GeV, and three radial
 bins in the neutrino interaction vertex. A $\chi^2$ analysis is performed,
 using the final NOMAD numbers
 on the observed and predicted electron-to-muon ratio, including statistical
 errors as well as the full error matrix describing systematic uncertainties
 and uncertainty correlations over different bins \cite{valuev}.
 In predicting the effect of $\nu_{\mu}\to\nu_e$ oscillations under
 any mass and mixing hypothesis, the contribution to $R_{e\mu}$ from
 oscillations with full mixing and $\Delta m^2=5,000\ \hbox{eV}^2$ 
 expected in NOMAD after all cuts is used \cite{valuev}.
 Energy resolution and baseline smearing effects
 (due to the long decay region) are taken into account \cite{nomad}. \\
%
%
\indent In Fig.~\ref{fig:fig8}, we show our calculations of the 90\% CL upper limits
 on oscillations as a function of $\Delta m^2$ for the six
 NSBL experiments considered here, as well as the 90\% CL allowed region
 for LSND. The $(\chi^2)_{\hbox{\tiny min}}$ values as a function of
 $\Delta m^2$ for all of the experiments are also shown.
 All the solid curves shown are obtained from the simplified
 analysis described here, and compare well with the published results
 \cite{Declais:1994su,Stockdale:1984cg,Dydak:1983zq,Apollonio:2002gd,
Armbruster:2002mp,lsnd,nomad}. \\
\indent The LSND region obtained in our analysis of
 DAR neutrinos is slightly shifted to the right compared to
 the final LSND area, shown in Fig.~\ref{fig:fig8}f as a dashed line,
 reflecting the difference in the two datasets. More detailed
 LSND DAR analyses give results in rough agreement with our
 allowed region \cite{Church:2002tc,Maltoni:2002xd}. \\
\indent The $(\chi^2)_{\hbox{\tiny min}}$ values obtained for the
 Bugey and CDHS experiments as a function of $\Delta m^2$ give
 details that might seem surprising, at first.
 Slightly better fits to the data are obtained under a
 neutrino oscillations hypothesis, as opposed to the no oscillations one.
 Therefore, we add a final comment to explain the results of these fits. \\
\indent The Bugey fit is driven by the data at the shortest
 baseline, 15 m, where the statistical errors on the observed
 positron spectrum from $\bar{\nu}_e$ interactions are the smallest.
 As explained in Ref.\cite{Declais:1994su},
 systematic uncertainties are taken into account by allowing for
 linear deformations, as a function of positron energy, of the
 ratio of observed to predicted positron yields. The values of
 $(\chi^2_{\hbox{\tiny Bugey}})_{\hbox{\tiny min}}$ as a function
 of $\Delta m^2$ are explained by the fact that, for certain
 $\Delta m^2$ values, an oscillatory fit to the 15 m
 positron spectrum ratio describes the data marginally better than any straight
 line. Our best-fit oscillation hypothesis to Bugey data only is
 $\Delta m^2=0.92\ \hbox{eV}^2$, $\sin^2 2\theta_{ee}=0.05$. \\
\indent For CDHS, the $(\chi^2_{\hbox{\tiny CDHS}})_{\hbox{\tiny min}}$
 curve in Fig.~\ref{fig:fig8}d  has a minimum at
 $\Delta m^2\simeq 20-30\ \hbox{eV}^2$.
 This minimum is due to the fact that the far/near $\nu_{\mu}$ rate ratio,
 corrected for the baseline and detector mass differences between the two
 detectors (as well as other minor effects), is measured to be slightly
 greater than one \cite{Dydak:1983zq}:
 $R_{\hbox{\tiny corr}}=1.044\pm 0.023\pm 0.025$. This
 marginal deviation from one causes the fit procedure to prefer
 more $\nu_{\mu}$ disappearance by oscillations in the near than in the
 far detector. Given the average $\nu_{\mu}$ energy (3.2 GeV) and
 pathlength (130 m) for neutrinos interacting in the CDHS near detector,
 this condition is satisfied in the $\Delta m^2=20-30\ \hbox{eV}^2$ range.
 Our best-fit oscillation hypothesis to CDHS data only is
 $\Delta m^2=24\ \hbox{eV}^2$, $\sin^2 2\theta_{\mu\mu}=0.29$. \\



\end{document}